\documentclass[journal]{IEEEtran}

\ifCLASSINFOpdf
\else
\fi

\usepackage{setspace}
\usepackage{bbm}
\usepackage[cmex10]{amsmath}
\usepackage{amssymb}
\usepackage{cite}
\usepackage{graphicx}
\usepackage{array,color}
\usepackage{amsmath}
\allowdisplaybreaks
\usepackage{stfloats}
\usepackage{graphicx}
\usepackage{epstopdf}
\usepackage{subfigure}
\usepackage{tabularx}
\usepackage{epsfig,epsf,color,balance,cite}
\usepackage{algorithmic}
\usepackage{algorithm}
\usepackage{url}
\usepackage{bm}
\usepackage{multirow}

\usepackage{amsthm}

\ifodd 1
\usepackage{soul}
\usepackage{color}
\setstcolor{red}

\newcommand{\del}[1]{\st{#1}} 

\newcommand{\com}[1]{\textbf{\color{red} (COMMENT: #1)}} 
\newcommand{\response}[1]{\textbf{\color{green} (RESPONSE: #1)}} 
\else

\newcommand{\del}[1]{}

\newcommand{\com}[1]{}
\newcommand{\comg}[1]{}
\newcommand{\response}[1]{}
\fi

\title{\huge {Intelligent Reflecting Surface-Aided Electromagnetic Stealth Against Radar Detection}}
\author{ 
	Beixiong Zheng,~\IEEEmembership{Senior Member,~IEEE}, Xue Xiong, Jie Tang,~\IEEEmembership{Senior Member,~IEEE}, \\
	and Rui Zhang,~\IEEEmembership{Fellow,~IEEE} 
	
		\thanks{
			B. Zheng and X. Xiong are with the School of Microelectronics, South China University
			of Technology, Guangzhou 511442, China (e-mail: bxzheng@scut.edu.cn; ftxuexiong@mail.scut.edu.cn).
			
			J. Tang is with the School of Electronic and Information Engineering, South China University
			of Technology, Guangzhou 510640, China (e-mail: eejtang@scut.edu.cn).
			
			R. Zhang is with the School of Science and Engineering, Shenzhen Research Institute of Big Data, The Chinese University of
			Hong Kong, Shenzhen, Guangdong 518172, China (e-mail: rzhang@cuhk.edu.cn). He is also with the Department of Electrical
			and Computer Engineering, National University of Singapore, Singapore 117583 (e-mail: elezhang@nus.edu.sg).
	
		}
}

\begin{document}
\maketitle
\begin{abstract}
While traditional electromagnetic stealth materials/metasurfaces can render a target virtually invisible to some extent, they lack flexibility and adaptability, and can only operate within a limited frequency and angle/direction range, making it challenging to ensure the expected stealth performance.
In view of this, we propose in this paper a new intelligent reflecting surface (IRS)-aided electromagnetic stealth system mounted on targets to evade radar detection, by utilizing the tunable passive reflecting elements of IRS to achieve flexible and adaptive electromagnetic stealth in a cost-effective manner. Specifically, we optimize the IRS's reflection at the target to minimize the sum received signal power of all adversary radars.
We first address the IRS's reflection optimization problem using the Lagrange multiplier method and derive a semi-closed-form optimal solution for the single-radar setup, which is then generalized to the multi-radar case. To meet real-time processing requirements, we further propose low-complexity closed-form solutions based on the reverse alignment/cancellation and minimum mean-square error (MMSE) criteria for the single-radar and multi-radar cases, respectively.
Additionally, we propose practical low-complexity estimation schemes at the target to acquire angle-of-arrival (AoA) and/or path gain information via a small number of receive sensing devices.
Simulation results validate the performance advantages of our proposed IRS-aided electromagnetic stealth system with the proposed IRS reflection designs.

\end{abstract}
\IEEEpeerreviewmaketitle

\section{Introduction}
The continuous evolution of radar technology has led to significant advancement, which has enhanced its performance drastically in contemporary military operations and civilian applications \cite{westra2009radar}. On the other hand, a critical challenge exists in preventing aircraft from being detected by adversary radars.
To achieve this, electromagnetic stealth technology has been widely adopted \cite{zikidis2014low,rao2002integrated,yuan2011properties}, which aims to minimize the reflection of electromagnetic waves by an aircraft in a radar system, thereby rendering the aircraft virtually invisible.
Extensive efforts have been devoted to the development of electromagnetic stealth materials to achieve superior stealth performance and make a target stealthier \cite{zikidis2014low,rao2002integrated,yuan2011properties}. Generally, electromagnetic stealth materials can be categorized into two types: 1) reflective stealth, which reduces the radar cross section (RCS) by reflecting the adversary radar's incident waves in various directions; and 2) absorptive stealth, which absorbs external electromagnetic waves \cite{vinoy1996radar,feng2006electromagnetic,ahmad2019stealth,micheli2014synthesis,bai2015reflections,wang2022structural}. 
Emerging electromagnetic stealth materials based on carbon/plasma/geographer \cite{ahmad2019stealth,micheli2014synthesis} or their multi-layer structures \cite{bai2015reflections,wang2022structural}, have been proposed to achieve relatively high absorbing efficiency.
However, they do not yield perfect stealth performance and only achieve a certain reduction in the RCS value of a target, especially in highly dynamic military scenarios.
Moreover, certain types of electromagnetic stealth materials can achieve expected stealth performance only for specific angles and frequencies of the incident electromagnetic waves. Emerging electromagnetic stealth metasurfaces, which arrange sub-wavelength elements with specific geometries in either periodic or non-periodic configurations, enable controlled electromagnetic wave manipulation for improved absorption characteristics \cite{liu2016novel, alaee2017theory}. 
However, designing specific geometric metasurfaces poses challenges related to working frequency, bandwidth, and performance.
It is noted that most existing electromagnetic stealth materials/metasurfaces lack flexibility and adaptability, and have high implementation complexity, especially in cases involving high-speed stealth targets and rapidly changing adversary radar modes. For instance, the high mobility of a target and rapid operational radar mode changes can result in quick alterations in detection frequency and incident wave angles, rendering electromagnetic stealth materials inapplicable to fast-changing environments. As such, besides relying solely on electromagnetic stealth materials/metasurfaces with fixed properties once fabricated, it becomes imperative to develop intelligent and adaptive electromagnetic stealth system integrated with reconfigurable/controllable metasurfaces to cope with increasingly sophisticated anti-radar detection scenarios.


Recent advances in micro electromechanical systems (MEMS) and digitally-controlled metasurfaces \cite{cui2014coding,liaskos2018new,liu2018programmable} have led to the development of a promising new technology known as  intelligent reflecting surface (IRS), or its equivalents such as reconfigurable intelligent surface (RIS) \cite{wu2021intelligent,qingqing2019towards,Renzo2019Smart,zheng2021survey}.
Essentially, IRS is a large electromagnetic metasurface composed of numerous low-cost passive reflecting elements, each of which can be digitally controlled to induce an independent amplitude change and/or phase shift to the incident signal, thereby collaboratively redirecting electromagnetic waves into desirable directions.
By dynamically configuring these elements based on system requirements, the IRS can proactively create favorable channel conditions and perform various functions, such as enhancing desired signal power, suppressing undesired interference, and refining channel statistics \cite{wu2021intelligent,zheng2021survey}.
One of the key advantages of the IRS is its reconfigurability in real time, which provides a new solution to cope with time-varying and highly dynamic wireless environments.
Unlike traditional active arrays equipped at the base station (BS) or relay, IRS only passively reflects the ambient radio signal without the need for any radio-frequency (RF) chains for signal processing or amplification, thus significantly reducing power consumption and hardware cost. Furthermore, IRS can be fabricated with a low profile, lightweight, and conformal geometry, which facilitates its flexible and dense deployment in future wireless communication and sensing systems. These characteristics make the IRS a vital component in the evolution of wireless communication infrastructure, capable of meeting the demands of increasingly sophisticated electromagnetic environments. As such, IRS has been extensively studied and incorporated into various communication and sensing systems, such as orthogonal frequency division multiplexing (OFDM), \cite{zheng2019intelligent,zheng2020intelligent,yang2019intelligent}, relaying communication \cite{zheng2021irs,Yildirim2021Hybrid,Abdullah2021Optimization}, multiple access \cite{Zheng2020IRSNOMA,Guo2021Intelligent,Zuo2021Reconfigurable}, wireless sensing \cite{Lin2022Sensing,buzzi2022foundations,Shao2022Target,wang2023target,Wang2022Joint,Hu2022Reconfigurable},
among others.

Most existing research has primarily focused on IRS-aided communication to enhance wireless communication capacity and reliability, or IRS-aided sensing to improve target sensing accuracy.  
For example, the joint active and passive beamforming optimization problem has been extensively investigated in single/double/multi-IRS systems \cite{Wu2019TWC,Zheng2020DoubleIRS,zheng2020efficient,mei2021intelligent,huang2021Multi-Hop,zheng2022intelligent}, multi-antenna communications \cite{wei2021channel,huang2020reconfigurable,zhang2019capacity}, and multi-cell networks \cite{Pan2020Multicell,Xie2021Max,Luo2021Reconfigurable}. 
More recently, initial works have introduced the concept of IRS-aided sensing to enhance radar detection \cite{buzzi2022foundations,Shao2022Target,wang2023target}. 
 In this context, IRS manipulates the signals coming from the radar transmitter/target and reflects them towards the target/radar receiver, thus enhancing radar sensing capabilities.
In addition to IRS's directional signal enhancement for communication/sensing, another important function of IRS is to achieve directional signal suppression by destructively combining signals reflected by IRS and propagated through other paths. 
While some works have considered the application of IRS for directional signal suppression (e.g., physical layer security \cite{Yu2022Robust,Shen2019Secrecy,Sheng2020Artificial}), the potential of IRS in electromagnetic stealth systems against radar detection remains largely unexplored. 
In contrast to traditional electromagnetic stealth technologies that rely on stealth materials with fixed characteristics and a limited working range in terms of frequency and angle/direction, IRS offers flexible and real-time control over incident electromagnetic waves and can work over a wide range of frequencies and angles/directions. This adaptability makes it more suitable to assist or complement imperfect electromagnetic stealth in highly dynamic wireless environments, thereby opening up new avenues for research in the field of anti-radar detection.

\begin{figure}[!t]
	\centering
	\includegraphics[width=3.5in]{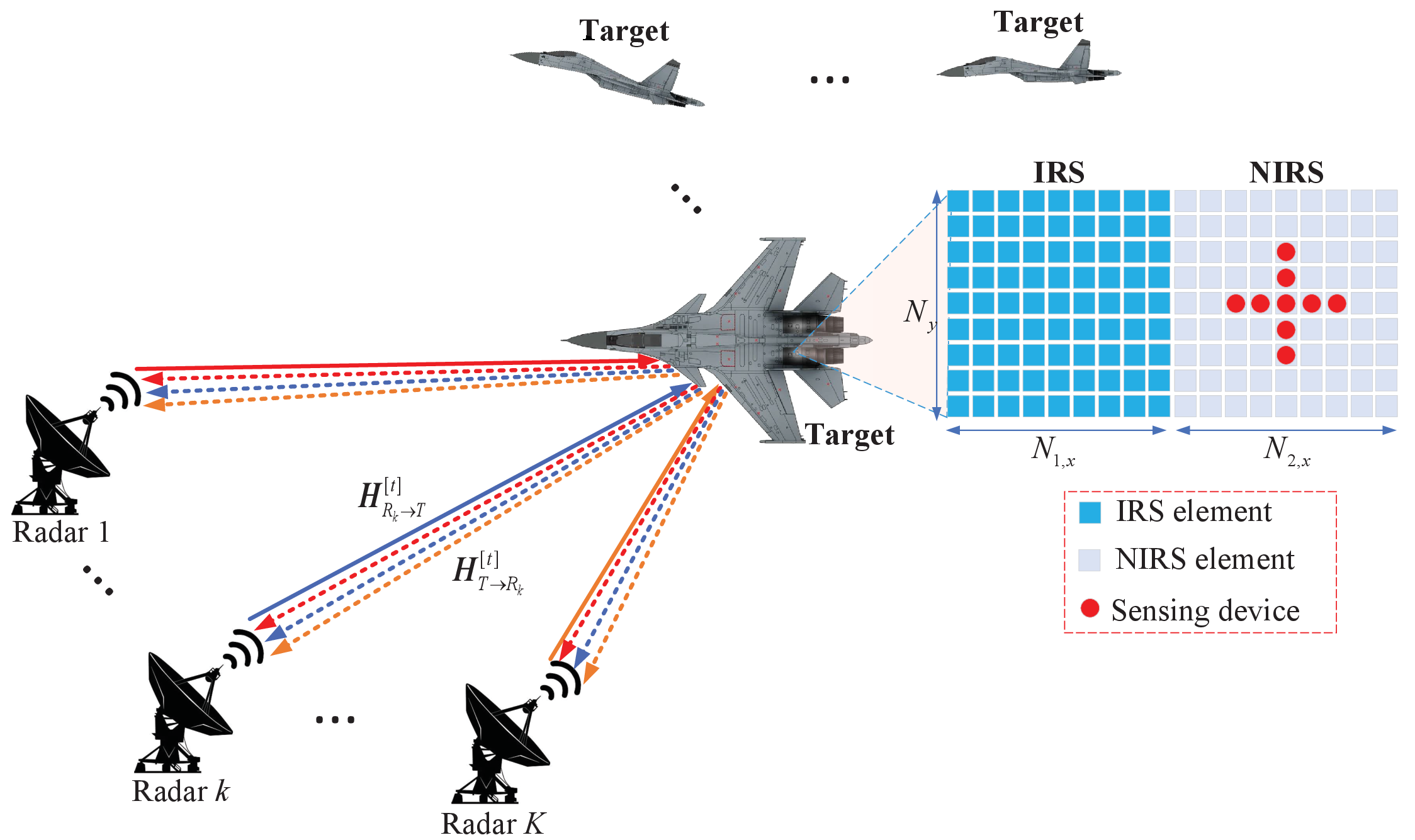}
	\setlength{\abovecaptionskip}{-3pt}
	\caption{Moving targets equipped with IRS-aided electromagnetic stealth against multiple mono-static radars.}
	\label{system}
\end{figure}
Motivated by the above, we study in this paper a new IRS-aided electromagnetic stealth system mounted on each moving target to evade detection from distributed radars as shown in Fig.~\ref{system}, where an IRS is deployed on the target to smartly and dynamically reduce the detection probability by the radars. In a general setup with co-existing echo-back and cross links among radars, we optimize the IRS's reflection at the target to minimize the sum received signal power of all adversary radars to evade detection, thereby achieving electromagnetic stealth.
Moreover, to address the angle-of-arrival (AoA) and path gain information required for the IRS's reflection design, we propose installing a cross-shaped array equipped with sensing devices at the target (see Fig.~\ref{system}). The main contributions of this paper are summarized as follows.
\begin{itemize}
	\item First, for the single-radar case, we optimally solve the IRS's reflection optimization problem via the Lagrange multiplier method, deriving a semi-closed-form optimal solution. To reduce complexity for real-time processing, we further propose a reverse alignment-based algorithm with a closed-form solution, which suppresses or even fully eliminates the signal reflected by the target. Additionally, by leveraging the Central Limit Theorem, we analytically determine the minimum number of IRS elements required to achieve full electromagnetic stealth in the proposed IRS-aided electromagnetic stealth system.
	\item Second, for the general multi-radar setup, we demonstrate that the IRS's reflection optimization problem is a convex quadratically constrained quadratic program (QCQP) that can be solved optimally. To provide insight into the structure of the optimal solution, we solve its Lagrangian dual problem and derive a semi-closed-form optimal solution. Moreover, to further reduce complexity for real-time processing, we propose leveraging the minimum mean-square error (MMSE) criterion for the IRS's reflection design to null or offset the signal reflected by the target over all echo-back and cross links among radars, and derive a closed-form MMSE solution.
	\item Third, to acquire the AoA and/or path gain information required by the IRS's reflection design, we develop practical low-complexity estimation schemes at the target, customized for both single-radar and multi-radar cases. Specifically, we exploit the structure of the cross-shaped sensing array (CSSA) to enable efficient estimation of AoA and path gain over the two-dimensional (2D) surface of the target, using only a small number of receive sensing devices.
	\item Finally, we provide extensive numerical results to validate the performance superiority of our proposed IRS-aided electromagnetic stealth system and the effectiveness of the proposed IRS's reflection designs. It is shown that the proposed IRS-aided electromagnetic stealth system significantly reduces and even eliminates the signal reflected by the target, thereby achieving satisfactory stealth performance. Besides, the effect of imperfect AoA information at the target on the radar performance is also investigated.
\end{itemize}

The rest of this paper is organized as follows. Section~\ref{sys} presents the system model and the
problem formulation for the IRS-aided electromagnetic stealth system. In Sections~\ref{Single-Radar} and \ref{Multi-Radar}, we propose efficient algorithms to solve the formulated problems for the IRS-aided electromagnetic stealth system under the single-radar and multi-radar setups, respectively. 
In Section~\ref{AoA}, we propose practical schemes to efficiently estimate the AoA and/or beamforming gain at the target for both single-radar and multi-radar cases.
Simulation results are presented in Section \ref{Sim} to evaluate the performance of the
proposed system and practical designs. Finally, conclusions are drawn in Section~\ref{conlusion}.

\emph{Notation:} 
Upper-case and lower-case boldface letters denote matrices and column vectors, respectively.
Superscripts ${\left(\cdot\right)}^{T}$, ${\left(\cdot\right)}^{H}$, ${\left(\cdot\right)}^{*}$, and ${\left(\cdot\right)}^{-1}$ stand for the transpose, Hermitian transpose, conjugate, and matrix inversion operations, respectively.
${\mathbb C}^{a\times b}$ denotes the space of ${a\times b}$ complex-valued matrices and
$n \mod{m}$ denotes the modulo operation which returns the remainder after division of $n$ by $m$.
For a complex-valued vector $\bm{x}$, $\lVert\bm{x}\rVert$ denotes its $\ell_2 $-norm,
$\angle (\bm{x} )$ returns the phase of each element in $\bm{x}$,
and ${\rm diag} (\bm{x})$ returns a diagonal matrix with the elements in $\bm{x}$ on its main diagonal.
$|\cdot|$ denotes the absolute value if applied to a complex-valued number or the cardinality if applied to a set.
${\cal O}(\cdot)$ stands for the standard big-O notation,
$\otimes$ denotes the Kronecker product, 
and $\odot$ denotes the Hadamard product.
${\bm I}$ and ${\bm 0}$ denote an identity matrix and an all-zero vector/matrix, respectively, with appropriate dimensions.
The distribution of a circularly symmetric complex Gaussian (CSCG) random vector with zero-mean and covariance matrix ${\bm \Sigma}$ is denoted by ${\mathcal N_c }({\bm 0}, {\bm \Sigma} )$; and $\sim$ stands for ``distributed as".

\section{System Model}\label{sys}

As shown in Fig. \ref{system}, we consider an IRS-aided electromagnetic stealth system mounted on a moving target (e.g., an aircraft) to evade detection from $K$ distributed radars, where the IRS and non-IRS (NIRS) such as electromagnetic wave absorbing materials are coated on the target's surface (TS) to reduce the detection probability by the adversary radars.\footnote{The results in this paper can be readily extended to the case with multiple moving targets, each equipped with an IRS-aided electromagnetic stealth system to independently evade detection from multiple adversary radars.}
Without loss of generality, we assume that each radar is mono-static (i.e., its transmitter and receiver are co-located) and equipped with a uniform
planar array (UPA) comprising $M\triangleq M_{x}\times M_{y}$ transmit/receive antennas, where $M_{x}$ and $M_{y}$ denote the numbers
of transmit/receive antennas along the $x$- and $y$-axes, respectively.
For the IRS-aided electromagnetic stealth system, both the IRS and NIRS are assumed to be UPAs, consisting of $N_{1}\triangleq N_{1,x}\times N_{1,y}$ and $N_{2}\triangleq N_{2,x}\times N_{2,y}$ passive elements, respectively, where $N_{1,x}$ ($N_{2,x}$) and $N_{1,y}$ ($N_{2,y}$) denote the numbers of elements at the IRS (NIRS) along the $x$- and $y$-axes, respectively.
In particular, the IRS and NIRS are combined to form a whole TS with a total number of $N=N_{1}+N_{2}\triangleq N_{x}\times N_{y}$ elements, where we assume $N_{x}=N_{1,x}+N_{2,x}$ and $N_{y}=N_{1,y}=N_{2,y}$ for simplicity.
The target is connected to a smart controller that is able to dynamically adjust IRS's reflection amplitudes and/or phase shifts in real time as well as coordinate the switching between two working modes, i.e., sensing mode for radar reconnaissance
and reflection mode for electromagnetic stealth.
Moreover, to enable sensing mode of the TS for radar reconnaissance, 
a cross-shaped array consisting of $L=L_{x}+ L_{y}-1$ receive sensing devices is embedded at the target, with $L_{x}$ and $L_{y}$ denoting the number of sensing devices along the $x$- and $y$-axes, respectively (see Fig. \ref{system}).

\subsection{Channel and Reflection Model}
As shown in Fig.~\ref{system}, due to the movement of the target, we denote  ${\bm H}_{R_k\rightarrow I}^{[t]} \in {\mathbb{C}^{N_1\times M }}$, ${\bm H}_{R_k\rightarrow N}^{[t]} \in {\mathbb{C}^{N_2\times M}}$, and ${\bm H}_{R_k\rightarrow T}^{[t]} \in {\mathbb{C}^{N\times M }}$
as the equivalent time-varying channels for the radar $k$$\rightarrow$IRS, radar $k$$\rightarrow$NIRS, and radar $k$$\rightarrow$TS links at time $t$, respectively.\footnote{For notational convenience, we use subscripts ``$R_k$", ``$I$", ``$N$", and ``$T$" to indicate radar $k$, IRS, NIRS, and TS, respectively.}
Due to the high altitude of aerial targets, the propagation channels from each mono-static radar to the IRS/NIRS/TS, i.e., $\left\{{\bm H}_{R_k\rightarrow I}^{[t]},  {\bm H}_{R_k\rightarrow N}^{[t]}, {\bm H}_{R_k\rightarrow T}^{[t]}\right\}$ can be characterized by the far-field line-of-sight (LoS) model with {\it parallel wavefronts}.
For convenience, we first define a one-dimensional (1D) steering vector function for a generic uniform linear array (ULA) as follows.
 \begin{align}
{\bm e}(\phi,{\bar N})\triangleq\left[1, e^{-j \pi \phi},\ldots, e^{-j \pi({\bar N}-1) \phi}\right]^T\in {\mathbb{C}^{{\bar N} \times 1 }}
\end{align}
where $j\triangleq \sqrt{-1}$ denotes the imaginary unit,
$\phi$ denotes the constant phase-shift difference between the signals at two adjacent antennas/elements, and
${\bar N}$ denotes the number of antennas/elements in the ULA.  
Due to the co-located IRS and NIRS on the TS, they share the same angle-of-arrival/departure (AoA/AoD) with each radar at a given time $t$, which can be denoted by the AoA/AoD pairs $(\vartheta_{R_k\rightarrow T}^{[t]}, \varphi_{R_k\rightarrow T}^{[t]})$ and $(\vartheta_{T\rightarrow R_k}^{[t]}, \varphi_{T\rightarrow R_k}^{[t]})$ at the TS and radar $k$, respectively, with $k=1,\ldots,K$. 
Accordingly, we let ${\bm a}_{T}(\vartheta_{R_k\rightarrow T}^{[t]}, \varphi_{R_k\rightarrow T}^{[t]})$ and
${\bm a}_{R_k}(\vartheta_{T\rightarrow R_k}^{[t]}, \varphi_{T\rightarrow R_k}^{[t]})$ denote the array response vectors of the TS and radar $k$, respectively, with $k=1,\ldots,K$.
Under the UPA model, each array response vector is expressed as the Kronecker product of two steering vector functions in the $x$-axis (horizontal) and $y$-axis (vertical) directions, respectively. 
Specifically, the array response vectors at the TS and radar $k$ are respectively expressed as
\begin{align}
\hspace{-0.2cm}{\bm a}_{T}(\vartheta_{R_k\rightarrow T}^{[t]}, &\varphi_{R_k\rightarrow T}^{[t]})\hspace{-0.1cm}=\hspace{-0.1cm}{\bm e}\hspace{-0.1cm}\left(\hspace{-0.1cm} \frac{2\Delta_e}{\lambda} \cos (\varphi_{R_k\rightarrow T}^{[t]}) \cos (\vartheta_{R_k\rightarrow T}^{[t]})  , N_{x} \hspace{-0.1cm}\right)\hspace{-0.1cm} \notag\\
&\otimes{\bm e}\left(\frac{2\Delta_e}{\lambda} \cos(\varphi_{R_k\rightarrow T}^{[t]})\sin (\vartheta_{R_k\rightarrow T}^{[t]}) , N_{y} \right) \\
\hspace{-0.2cm}{\bm a}_{R_k}(\vartheta_{T\rightarrow R_k}^{[t]}, &\varphi_{T\rightarrow R_k}^{[t]})\hspace{-0.1cm}=\hspace{-0.1cm}{\bm e}\hspace{-0.1cm}\left(\hspace{-0.1cm} \frac{2\Delta_a}{\lambda} \cos (\varphi_{T\rightarrow R_k}^{[t]}) \cos (\vartheta_{T\rightarrow R_k}^{[t]}) , M_{x} \hspace{-0.1cm}\right)\hspace{-0.1cm}  \notag\\
&\otimes{\bm e}\left(\frac{2\Delta_a}{\lambda} \cos(\varphi_{T\rightarrow R_k}^{[t]})\sin (\vartheta_{T\rightarrow R_k}^{[t]}) , M_{y} \right)
\end{align}
where $\lambda$ denotes the signal wavelength, $\Delta_e$ is the element spacing at the TS, and
$\Delta_a$ is the antenna spacing at each radar; while the  array response
vectors at the IRS and NIRS, denoted by ${\bm a}_{I}(\vartheta_{R_k\rightarrow T}^{[t]}, \varphi_{R_k\rightarrow T}^{[t]})$ and ${\bm a}_{N}(\vartheta_{R_k\rightarrow T}^{[t]}, \varphi_{R_k\rightarrow T}^{[t]})$, respectively, can be similarly defined by virtue of the same AoA/AoD pair of $(\vartheta_{R_k\rightarrow T}^{[t]}, \varphi_{R_k\rightarrow T}^{[t]})$ due to their co-location relationship.
Recall that the TS is concatenated by the IRS and NIRS as shown in Fig.~\ref{system} and the relationship among the array response vectors of the TS, IRS, and NIRS is characterized by
\begin{align}
\hspace{-0.2cm}\underbrace{{\bm a}_{T}^T\hspace{-0.05cm}(\hspace{-0.05cm}\vartheta_{R_k\rightarrow T}^{[t]},\hspace{-0.1cm} \varphi_{R_k\rightarrow T}^{[t]}\hspace{-0.05cm})\hspace{-0.05cm}}_{={\bm e}^T_{x}\otimes{\bm e}^T_{y}}\hspace{-0.1cm}=\hspace{-0.15cm}\Big[\hspace{-0.1cm}\underbrace{{\bm a}_{I}^T\hspace{-0.05cm}(\hspace{-0.05cm}\vartheta_{R_k\rightarrow T}^{[t]},\hspace{-0.1cm} \varphi_{R_k\rightarrow T}^{[t]}\hspace{-0.05cm})\hspace{-0.05cm}}_{={\bm e}^T_{1,x}\otimes{\bm e}^T_{y}},  \underbrace{{\bm a}_{N}^T\hspace{-0.05cm}(\hspace{-0.05cm}\vartheta_{R_k\rightarrow T}^{[t]},\hspace{-0.1cm} \varphi_{R_k\rightarrow T}^{[t]}\hspace{-0.05cm})\hspace{-0.05cm}}_{={\bm e}^T_{2,x}\otimes{\bm e}^T_{y}}\hspace{-0.1cm}\Big]\hspace{-0.2cm}\notag
\end{align}
which is obtained by using the property of ${\bm e}^T_{x}\otimes{\bm e}^T_{y}=\left[{\bm e}^T_{1,x}, {\bm e}^T_{2,x}\right]\otimes{\bm e}^T_{y}=\left[{\bm e}^T_{1,x}\otimes{\bm e}^T_{y}, {\bm e}^T_{2,x}\otimes{\bm e}^T_{y}\right]$.
Accordingly, the real-time far-field LoS channels between any two nodes (represented by $X$ and $Y$ for notational simplicity) are modeled as the outer product of array responses at their two sides, i.e., 
\begin{align}\label{Far_LoS}
\hspace{-0.2cm}{{\bm H}}_{Y\rightarrow X}^{[t]}=\rho_{Y-X}^{[t]}   {\bm a}_X( \vartheta_{Y\rightarrow X}^{[t]}, \varphi_{Y\rightarrow X}^{[t]})  {\bm a}^T_Y( \vartheta_{X\rightarrow Y}^{[t]}, \varphi_{X\rightarrow Y}^{[t]})\hspace{-0.2cm} 
\end{align}
with $X \in \{I, N, T\}$ and $Y \in \{R_k\}_{k=1}^K$,
where $\rho_{Y-X}^{[t]}\triangleq\frac{\sqrt{\alpha}}{d_{X-Y}^{[t]}} e^{\frac{-j 2 \pi  }{\lambda}d_{X-Y}^{[t]}}$ is the corresponding complex-valued path gain between them at time $t$ with
$\alpha$ being the reference path gain at the distance of 1 meter (m) and 
$d_{X-Y}^{[t]} $ denoting the real-time propagation distance between nodes $X$ and $Y$.
Moreover, due to the co-location of the IRS and NIRS on the TS, we can obtain the channel relationship among the TS, IRS, and NIRS as 
$
{\bm H}_{R_k\rightarrow T}^{[t]}=\begin{bmatrix} {\bm H}_{R_k\rightarrow I}^{[t]}\\{\bm H}_{R_k\rightarrow N}^{[t]} \end{bmatrix}
$.
In addition, under the far-field LoS channel model, the channel reciprocity holds for each link in forward and reverse directions, and thus we have
$
{{\bm H}}_{X\rightarrow Y}^{[t]}=\left({{\bm H}}_{Y\rightarrow X}^{[t]}\right)^T
$
with $X \in \{I, N, T\}$ and $Y \in \{R_k\}_{k=1}^K$.

Given practical constraints such as physical characteristics, absorbed frequency range, and structural layout, existing coating materials on the target cannot fully absorb all electromagnetic waves radiated from multiple distributed radars. To account for this imperfect absorption as well as non-absorbing materials coated on the target, we introduce the term ``absorbing efficiency" to quantify the effectiveness of the NIRS elements in absorbing electromagnetic waves from adversary radars, which is denoted by $\zeta_n\in [0, 1], \forall n=1,\ldots,N_2$. 
Accordingly, we let ${\bm \phi}\triangleq  [\sqrt{1-\zeta_1} e^{j\phi_{1}},\ldots,\sqrt{1-\zeta_{N_2}} e^{j\phi_{N_2}}]^T$ denote the equivalent (non-tunable) reflection coefficients of the NIRS, with $\{\phi_{n}\}_{n=1}^{N_2}$ being the (non-tunable) phase shifts determined by the NIRS property.\footnote{In practice, the equivalent (non-tunable) reflection coefficients of the NIRS can be obtained through an offline process that involves precise modeling and extensive measurement.}
On the other hand, we let ${\bm \theta}^{[t]}\triangleq\left[{\theta_{1}^{[t]}},\ldots,{\theta_{N_1}^{[t]}}\right]^T=\left[\beta_{1}^{[t]} e^{j\psi_{1}^{[t]}},\ldots, \beta_{N_1}^{[t]} e^{j\psi_{N_1}^{[t]}}\right]^T$ denote the equivalent (tunable) reflection coefficients of the IRS, where $\beta_{n}^{[t]}\in [0, \beta_\text{max}]$ and $\psi_n^{[t]} \in [0, 2\pi)$ with $n=1,2,\ldots,N_1$ are the reflection amplitude and phase shift of the $n$-th element at time $t$, respectively, with $\beta_\text{max}\le 1$ being the maximum reflection amplitude of each IRS element, i.e., $\left|{\theta_{n}^{[t]}}\right|\le \beta_\text{max}, \forall n=1,2,\ldots,N_1$.
For simplicity, we let ${\bar{\bm \theta}}^{[t]}=\left[ \left({\bm \theta}^{[t]}\right)^T, {\bm \phi}^T\right]^T$ denote the concatenated reflection vector of the TS.

\subsection{Signal Model}
During the radar detection process, we assume that each mono-static radar transmits one coherent burst of $L$ non-consecutive radar pulses with a constant pulse repetition interval (PRI), denoted as $T_p$, to detect the moving target. 
The duration over which all these signals are reflected by the target and received by each radar is referred to as the coherent-processing interval (CPI), denoted by $T_c$, which is equal to $LT_p$.
The pulse duration of each radar is denoted as $t_p$, with $t_p<T_p$.
Moreover, we assume that the target location, as well as the PRI and pulse duration of each radar, remain unchanged during each CPI, while they may change from one CPI to another. The radar pulse waveform of each radar during each PRI is given by
\begin{align}\label{waveform}
{x}_k(t)=\left\{
\begin{aligned}
&\sqrt{P} {\bar x}_k(t), &0\le t\le t_p\\
&0, &t_p< t\le T_p
\end{aligned}
\right., ~~~\forall k=1,\ldots,K
\end{align}
where $P$ denotes the radar transmit power and ${\bar x}_k(t)$ is the corresponding
radar pulses with normalized power, i.e., $\frac{1}{T_p}\int_{0}^{t_p} |{\bar x}_k(t)|^2 {\rm d}t =1$.

Based on the above, the radar probing signals echoed back from the target and then received by each radar at time $t$ can be expressed as
\begin{align}\label{rec_sig}
\hspace{-0.2cm}y_k(t)\hspace{-0.1cm}=\hspace{-0.1cm} {\bm w}_k^T {\bm H}_{T\rightarrow R_k}^{[t]} \text{diag}\hspace{-0.1cm}\left(\hspace{-0.1cm} {\bar{\bm \theta}}^{[t]}\hspace{-0.1cm}\right) \hspace{-0.15cm}
	\left(\hspace{-0.05cm}\sum_{j=1}^{K} {\bm H}_{R_j\rightarrow T}^{[t]} {\bm w}_j{x}_j(t)\hspace{-0.1cm}\right) \hspace{-0.15cm}+\hspace{-0.1cm}n_k(t)\hspace{-0.2cm}
\end{align}
with $k=1,\ldots,K$, where ${\bm w}_k$ and ${\bm w}_k^T$ stand for the transmit and matching receive beamformers, respectively, and
$n_k(t)\sim {\mathcal N_c }(0, \sigma^2)$ is the zero-mean additive white Gaussian noise (AWGN) with variance of $\sigma^2$  at each mono-static radar.

Note that radar detects a target only when it receives an adequate amount of power back from the target.
In radar detection, performance metrics such as target presence detection and AoA estimation heavily rely on the received signal power. A larger reflected signal power from the target typically results in a higher signal-to-noise ratio (SNR), leading to improved detection probability and/or more accurate AoA estimation.  Conversely, the target aims to minimize or eliminate the reflected signal power to evade detection. In this context, we use the radars' received signal power as the performance metric to evaluate the effectiveness of IRS-aided electromagnetic stealth against radar detection.
We assume the target location remains approximately constant during each CPI, resulting in negligible variations in channels and geometry-related parameters (e.g., propagation distances and AoA/AoD) between the target and radars. 
As such, we omit the time index $[t]$ for brevity without causing any confusion in the following.
According to \eqref{rec_sig}, the received signal power at radar $k$ over one PRI is given by
\begin{align}\label{power}
\hspace{-0.2cm}{\bar P}_k=&\sum_{j=1}^{K}\left|{\bm w}_k^T {\bm H}_{R_k\rightarrow T}^T \text{diag}\left( {\bar{\bm \theta}}\right) 
 {\bm H}_{R_j\rightarrow T} {\bm w}_j\right|^2\frac{1}{T_p}\int_{0}^{T_p} |x_j(t)|^2 {\rm d}t\notag\\
 =&P\sum_{j=1}^{K} \Big| \underbrace{\rho_{{R_k}-T}
 {\bm w}_k^T{\bm a}_{R_k}( \vartheta_{T\rightarrow R_k}, \varphi_{T\rightarrow R_k}) }_{G_{R,k}: \rm receive~beamforming~gain} \notag\\
 &\underbrace{{\bm a}^T_{T}( \vartheta_{R_k\rightarrow T}, \varphi_{R_k\rightarrow T})
 \text{diag}\left( {\bar{\bm \theta}}\right)
 {\bm a}_{T}( \vartheta_{R_j\rightarrow T}, \varphi_{R_j\rightarrow T})}_{R_{k,j}\left( {\bar{\bm \theta}}\right):  \rm reflection~gain}  \notag\\
 & \underbrace{\rho_{{R_j}-T}{\bm a}^T_{R_j}( \vartheta_{T\rightarrow R_j}, \varphi_{T\rightarrow R_j}) 
 {\bm w}_j}_{G_{T,j}:\rm transmit~beamforming~gain} \Big|^2 \notag\\
=&P\sum_{j=1}^{K} \Big|G_{R,k} \Big|^2 \Big|G_{T,j} \Big|^2 \Big| {\bar{\bm u}}_{k,j}^H {\bar{\bm \theta}} \Big|^2
\end{align}
where $G_{T,k}$ and
$G_{R,k}$ denote the complex-valued transmit and receive beamforming gains at radar $k$, respectively,
$R_{k,j}\left( {\bar{\bm \theta}}\right)={\bar{\bm u}}_{k,j}^H {\bar{\bm \theta}}$ denotes the complex-valued reflection gain at the TS that depends on the IRS reflection,
and
\begin{align}\label{cascaded}
\hspace{-0.3cm}{\bar{\bm u}}_{k,j}^H=&{\bm a}_{T}^T( \vartheta_{R_k\rightarrow T}, \varphi_{R_k\rightarrow T})
\odot
{\bm a}_{T}^T( \vartheta_{R_j\rightarrow T}, \varphi_{R_j\rightarrow T})\notag\\
=&\big[\underbrace{{\bm a}_I^T( \vartheta_{R_k\rightarrow T}, \varphi_{R_k\rightarrow T})
\odot
{\bm a}_I^T( \vartheta_{R_j\rightarrow T}, \varphi_{R_j\rightarrow T})}_{{\bm u}_{k,j}^H:\rm cascaded~array~response~vector~at~IRS},\notag\\ &\underbrace{{\bm a}_N^T( \vartheta_{R_k\rightarrow T}, \varphi_{R_k\rightarrow T})
\odot
{\bm a}_N^T( \vartheta_{R_j\rightarrow T}, \varphi_{R_j\rightarrow T})}_{{\tilde{\bm u}}_{k,j}^H:\rm cascaded~array~response~vector~at~NIRS}\big]\hspace{-0.3cm}
\end{align}
is the cascaded array response vector at the TS.
 Based on \eqref{power} and \eqref{cascaded}, the sum received signal power at multiple radars is given by
\begin{align}\label{sumpower}
{\bar P}_{\rm sum}&=\sum_{k=1}^{K}{\bar P}_k=P\sum_{k=1}^{K}\sum_{j=1}^{K} \Big|G_{R,k} \Big|^2 \Big|G_{T,j} \Big|^2 \Big| {\bar{\bm u}}_{k,j}^H {\bar{\bm \theta}} \Big|^2\notag\\
&=P\sum_{k=1}^{K}\sum_{j=1}^{K} \Big|G_{R,k} \Big|^2 \Big|G_{T,j} \Big|^2 \Big| {\bm u}_{k,j}^H {\bm \theta} + {\tilde{\bm u}}_{k,j}^H{\bm \phi} \Big|^2.
\end{align}
\subsection{Problem Formulation}
In this paper, we aim to minimize the sum received signal power of all adversary radars to evade detection (i.e., to achieve electromagnetic stealth) for the target, by optimizing the IRS's reflection vector ${\bm \theta}$ to synergize with the NIRS.
Accordingly, the problem is formulated as
\begin{align}
\text{(P1):}
& \underset{ {\bm \theta} }{\text{min}}
& &  \sum_{k=1}^{K}\sum_{j=1}^{K} \Big|G_{R,k} \Big|^2 \Big|G_{T,j} \Big|^2 \Big| {\bm u}_{k,j}^H {\bm \theta} + {\tilde{\bm u}}_{k,j}^H{\bm \phi} \Big|^2 \label{obj_P1} \\
& \text{s.t.} & &|{\theta}_{n}|\le \beta_\text{max}, \forall n=1,\ldots,N_1 \label{con1_P1}.
\end{align}
It can be verified that problem (P1) is a convex QCQP, which can thus be optimally solved by existing convex optimization solvers such as CVX \cite{grant2014cvx}. However, this numerical solution offers limited insight into the structure of its optimal solution and has a relatively high computational complexity, which may not be suitable for real-time processing in time-critical applications. To address this issue, we propose low-complexity algorithms with (semi)-closed-form solutions to optimally solve (P1) in the single-radar case at first, which is then generalized to the multi-radar case, thereby providing a more efficient and insightful approach to solving the problem.

\section{IRS Reflection Design in Single-Radar Case}\label{Single-Radar}
In this section, we consider the single-radar setup, i.e., $K = 1$, to draw essential and useful insights into the optimal reflection design for minimizing the received signal power at one mono-static radar without cross link. 
For brevity, the user index $k$ or $j$ can be dropped in this section. As such, with
the absence of cross link, problem (P1) can be simplified as follows (with constant/irrelevant terms omitted for brevity).
\begin{align}
\text{(P2):}
& \underset{ {\bm \theta} }{\text{min}}
& &  \Big| {\bm u}^H {\bm \theta} + C \Big|^2 \label{obj_P2} \\
& \text{s.t.} & &|{\theta}_{n}|\le\beta_\text{max}, \forall n=1,\ldots,N_1 \label{con1_P2}
\end{align}
where $C\triangleq{\tilde{\bm u}}^H{\bm \phi}$ is a complex-valued reflection gain at the NIRS\footnote{The complex-valued reflection gain $C$, which is dependent on both the AoA/AoD and reflection coefficients at the NIRS, can also be practically acquired from an offline database through extensive measurement.} and ${\bm u}\triangleq\left[u_{1},\ldots,u_{N_1}\right]^T$ is cascaded array response vector at the IRS as defined in \eqref{cascaded}. 
It is noted that solving problem (P2) at the target (IRS-aided electromagnetic stealth system) only requires the AoA information from the radar to target, i.e., $( \vartheta_{R\rightarrow T}, \varphi_{R\rightarrow T})$, which can be effectively estimated via the receive sensing devices embedded at the target, as will be elaborated in Section \ref{AoA}.
In the following, we address problem (P2) by providing low-complexity algorithms with (semi)-closed-form solutions.

\subsection{Lagrange Multiplier-Based Solution}\label{Lagrange_single}
When a problem is formulated as a QCQP problem, as in the case of (P2), it is often beneficial to explore its dual to reveal the structure of its optimal solution. This approach is illustrated by demonstrating that the Lagrangian dual of problem (P2) has a meaningful engineering interpretation. To facilitate problem-solving, problem (P2) can be reformulated as
\begin{align}
\text{(P2.1):}
& \underset{ {{\bm \theta}} }{\text{min}}
& &  {\bm \theta}^H {\bm u} {\bm u}^H {\bm \theta}+ C{\bm \theta}^H {\bm u} +C^*{\bm u}^H{\bm \theta}+ |C|^2\label{obj_P2.1} \\
& \text{s.t.} & & {\theta}_{n}^*{\theta}_n\le\beta^2_\text{max}, \forall n=1,\ldots,N_1. \label{con1_P2.1}
\end{align}
Accordingly, the Lagrange multiplier function for problem (P2.1) is defined as
\begin{align}
\hspace{-0.2cm}&{\cal L}\hspace{-0.07cm}\left({\bm \theta}, { {\bm \lambda}}\right)\hspace{-0.07cm}\triangleq\hspace{-0.07cm}  {\bm \theta}^H\hspace{-0.05cm} {\bm u} {\bm u}^H\hspace{-0.05cm} {\bm \theta}\hspace{-0.07cm}+\hspace{-0.07cm} C{\bm \theta}^H\hspace{-0.05cm} {\bm u} \hspace{-0.07cm}+\hspace{-0.07cm}C^*\hspace{-0.05cm}{\bm u}^H\hspace{-0.05cm}{\bm \theta}\hspace{-0.07cm}+\hspace{-0.07cm} |C|^2\hspace{-0.07cm}+\hspace{-0.07cm}\sum_{n=1}^{N_1}\hspace{-0.07cm} \lambda_n\left( {\theta}_{n}^*{\theta}_n \hspace{-0.07cm}-\hspace{-0.07cm}\beta^2_\text{max} \right)
\notag\\
\hspace{-0.2cm}&\hspace{-0.12cm}=\hspace{-0.12cm}
 {\bm \theta}^H\hspace{-0.15cm} \left(\hspace{-0.07cm}{\bm u} {\bm u}^H\hspace{-0.1cm}+\hspace{-0.07cm}\text{diag}\hspace{-0.07cm}\left({\bm \lambda}\hspace{-0.05cm}\right) \hspace{-0.05cm}\right)\hspace{-0.07cm} {\bm \theta}\hspace{-0.07cm}+\hspace{-0.07cm}C{\bm \theta}^H\hspace{-0.07cm} {\bm u} \hspace{-0.07cm}+\hspace{-0.07cm}C^*{\bm u}^H\hspace{-0.07cm}{\bm \theta}\hspace{-0.07cm}+\hspace{-0.07cm} |C|^2\hspace{-0.07cm}-\hspace{-0.07cm}\beta^2_\text{max}\hspace{-0.07cm}{\bm 1}^T\hspace{-0.07cm}{\bm \lambda}\hspace{-0.2cm}
\label{Lagrange2}
\end{align}
where ${\bm \lambda}\triangleq[\lambda_1, \ldots ,\lambda_{N_1}]^T$ is the vector of Lagrange multipliers with $\lambda_n\ge0, \forall n=1,\ldots,N_1$.
Then, we take the differentiation of ${\cal L}\left({\bm \theta}, {\bm \lambda}\right)$ with respect to ${\bm \theta}$, yielding
\begin{align}
\frac{\partial{\cal L}\left({\bm \theta}, {{\bm \lambda}}\right)}{\partial{\bm \theta}}&=\left({\bm u} {\bm u}^H + \text{diag}\left({{\bm \lambda}}\right) \right) {\bm \theta} + C{\bm u}.
\end{align}
 By letting $\frac{\partial{\cal L}\left({\bm \theta}, {{\bm \lambda}}\right)}{\partial{\bm \theta}}=0$, we arrive at a semi-closed-form solution:
\begin{align}\label{opt}
{\bm \theta}^{\star}=-C\left({\bm u} {\bm u}^H + \text{diag}\left({{\bm \lambda}}\right) \right)^{-1}{\bm u}.
\end{align}
Apparently, the optimal value of ${\bm \theta}$ is a function of the Lagrange multiplier vector ${\bm \lambda}$.
Since problem (P2.1) is a QCQP with zero-duality gap, we can obtain the dual
variables $\lambda_n\ge0, \forall n=1,\ldots,N_1$ through its dual program.
Accordingly, by substituting \eqref{opt} into \eqref{Lagrange2},
we can obtain the dual function as 
\begin{align}\label{dual}
g\left({\bm \lambda}\right)& = \underset{ {\bm \theta} }{\text{inf}} ~{\cal L}\left({\bm \theta}, {{\bm \lambda}}\right)\notag\\
&=
\left\{
\begin{aligned}
&|C|^2-\beta^2_\text{max}{\bm 1}^T{\bm \lambda}-|C|^2{\bm u}^H {\bar{\bm U}}^{-1}{\bm u}, {\bar{\bm U}} \succcurlyeq 0\\
&-\infty, \quad {\rm otherwise}
\end{aligned}
\right.
\end{align}
where ${\bar{\bm U}}={\bm u} {\bm u}^H+\text{diag}\left({\bm \lambda}\right)$.
Then, by exploiting the Schur complement, the dual problem can be expressed as an equivalent semidefinite
optimization problem, i.e., 
\begin{align}
\hspace{-0.25cm}\text{(P2.2):}
& \underset{ q, {\bm \lambda} }{\text{max}}
& &  q\label{obj_P2.2} \\
& \text{s.t.} & & 
\begin{bmatrix}
|C|^2-\beta^2_\text{max}{\bm 1}^T{\bm \lambda}\hspace{-0.1cm}-\hspace{-0.1cm}q&C^*{\bm u}^H\\C{\bm u}  &{\bm u} {\bm u}^H+\text{diag}\left({\bm \lambda}\right)
\end{bmatrix}\hspace{-0.1cm}\succcurlyeq\hspace{-0.1cm} 0\hspace{-0.1cm}\label{con1_P2.2}\\
& & & \lambda_n \ge 0, \forall n=1,\ldots,N_1\label{con2_P2.2}
\end{align}
which can be effectively solved via standard semidefinite program (SDP) or linear
matrix inequalities (LMI) optimization, with the complexity order of ${\cal O}(N_1^{4.5})$ \cite{Luo2010Semidefinite}.

\subsection{Reverse Alignment-Based Solution} \label{Alignment_single}
To achieve lower complexity for real-time application, we further propose a reverse alignment-based algorithm with a closed-form solution. 
Intuitively, by taking a closer look at the objective function in \eqref{obj_P2} of problem (P2), we find that ${\bm \theta}$ can be designed to mitigate or even fully eliminate the complex-valued reflection gain from the NIRS, i.e., $C$, via reverse alignment/cancellation. Specifically, depending on the number of IRS elements, we consider the following two cases.
\begin{itemize}
	\item {\bf Case 1}: $N_1 <\left\lceil \frac{|C|}{\beta_\text{max}} \right\rceil $. 
	In this case, the IRS reflection coefficients can be designed as
\begin{align}\label{case1}
	\theta_n=-\beta_\text{max} u_n\frac{ C}{|C|} , \qquad \forall n=1,\ldots, N_1.
\end{align}
By substituting \eqref{case1} into the objective function in \eqref{obj_P2} of problem (P2), we can obtain the corresponding objective value as
\begin{align}
\Big| N_1 \beta_\text{max}\frac{ C}{|C|}  - C \Big|^2=
(|C|-N_1\beta_\text{max})^2 
\end{align}
which can be regarded as the residual reflection power from the target.
	\item {\bf Case 2}: $N_1 \ge \left\lceil \frac{|C|}{\beta_\text{max}} \right\rceil$. In this case, 
	since $N_1 \beta_\text{max} \ge |C|$,
	the IRS reflection coefficients for the first $\left\lceil \frac{|C|}{\beta_\text{max}} \right\rceil$ elements can be similarly designed as \eqref{case1}, $\forall n=1,\ldots,  \left\lceil \frac{|C|}{\beta_\text{max}} \right\rceil$;
	while the remaining reflection coefficients, i.e., $\theta_n$ with $n=\left\lceil \frac{|C|}{\beta_\text{max}} \right\rceil+1,\ldots, N_1 $ can be easily designed to fully 
	eliminate the complex-valued residual reflection gain from the NIRS, i.e.,
	\begin{align}\label{remain_coeff}
	\sum_{n=\left\lceil \frac{|C|}{\beta_\text{max}} \right\rceil+1}^{N_1} u_n^* \theta_n=-\left(|C| \mod{\beta_\text{max}}\right)\frac{ C}{|C|}.
	\end{align}
	For the IRS's reflection design based on \eqref{case1} and \eqref{remain_coeff}, we have ${\bm u}^H {\bm \theta}-C=0$ and thus achieve the optimal objective value of $0$ in \eqref{obj_P2}, i.e., full electromagnetic stealth.
\end{itemize}
It is readily obtained that the reverse alignment-based solution only has a linear complexity, i.e., ${\cal O}(N_1)$, which is appealing to real-time application.
\subsection{Minimum Number of IRS Elements for Electromagnetic Stealth}
Next, we analyze the minimum number of IRS elements required to achieve full electromagnetic stealth in the proposed IRS-aided electromagnetic stealth system. Specifically, based on the results presented in Section \ref{Alignment_single}, for a given complex-valued reflection gain from the NIRS, i.e., $C$, 
 the required minimum number of IRS elements can be readily calculated as
$N_1^{\min}= \left\lceil \frac{|C|}{\beta_\text{max}} \right\rceil$.
However, from a practical perspective, it is more interesting to pre-determine the minimum number of IRS elements, using only the statistical information of the reflection gain from the NIRS, rather than its instantaneous information.
By leveraging the substantial number of NIRS elements and the Central Limit Theorem, the complex-valued reflection gain from the NIRS $C$ can be effectively approximated as a complex zero-mean Gaussian random variable, whose variance (average power) is given by
\begin{align}
\sigma^2_C&={\mathbb E}\{|C|^2\}={\mathbb E}\{{\tilde{\bm u}}^H{\bm \phi}{\bm \phi}^H {\tilde{\bm u}}\}\notag\\
&={\tilde{\bm u}}^H{\mathbb E}\{{\bm \phi}{\bm \phi}^H \} {\tilde{\bm u}}=(1-{\bar \zeta})N_2
\end{align}
where ${\bar \zeta}=\frac{1}{N_2}\sum_{n=1}^{N_2}\zeta_n$.
In addition, it can be readily inferred that the reflection gain from the NIRS follows the exponential distribution, i.e., $|C|^2 \sim \exp\left({1}/{\sigma^2_C}\right)$. 
Suppose $|C_1|^2, |C_2|^2, \ldots, |C_I|^2$ are independent and identically distributed (i.i.d.) exponential variables for different realizations. We let $|{\bar C}|^2=\max\left\{|C_1|^2, |C_2|^2, \ldots, |C_I|^2\right\}$ denote the maximum of independent exponential variables, whose statistical expectation is given by
\begin{align}
{\mathbb E}\{|{\bar C}|^2\}&={\mathbb E}\left\{ \max\left\{|C_1|^2, |C_2|^2, \ldots, |C_I|^2\right\} \right\}\notag\\
&=\sum_{i=1}^{I} \frac{\sigma^2_C}{i}=\sum_{i=1}^{I} \frac{(1-{\bar \zeta})N_2}{i}.
\end{align}
As such, to offset the maximum signal power reflected from the NIRS among different realizations, the required minimum number of IRS elements is thus calculated as
\begin{align}\label{N_1min}
N_1^{\min}= \left\lceil\frac{ \sqrt{{\mathbb E}\{|{\bar C}|^2\}} }{\beta_\text{max}}\right\rceil
=\left\lceil\sqrt{ \sum_{i=1}^{I}\frac{(1-{\bar \zeta})N_2}{i\beta_\text{max}^2}}\right\rceil.
\end{align}

\section{IRS Reflection Design in Multi-Radar Case}\label{Multi-Radar}
In this section, we consider the general multi-radar setup for the IRS-aided electromagnetic stealth system.
In the following, we propose two efficient algorithms to solve (P1) efficiently.
\subsection{Lagrange Multiplier-Based Solution}\label{Lagrange_multi}
Problem (P1) can be reformulated as
\begin{align}
\text{(P3):}
& \underset{ {\bm \theta} }{\text{min}}
& &  {\bm \theta}^H {\tilde{\bm U}} {\bm \theta}+ {\bm \theta}^H {\tilde{\bm v}} +{\tilde{\bm v}}^H{\bm \theta}+ {\tilde C}\label{obj_P3} \\
& \text{s.t.} & &{\theta}_{n}^*{\theta}_n\le\beta^2_\text{max}, \forall n=1,\ldots,N_1\label{con1_P3}
\end{align}
where
\begin{align}
{\tilde{\bm U}}&=\sum_{k=1}^{K}\sum_{j=1}^{K} \Big|{\tilde G}_{k,j} \Big|^2  {\bm u}_{k,j} {\bm u}_{k,j}^H \\
{\tilde{\bm v}}&=\sum_{k=1}^{K}\sum_{j=1}^{K} \Big|{\tilde G}_{k,j} \Big|^2  {\tilde{\bm u}}_{k,j}^H{\bm \phi}{\bm u}_{k,j}\\
{\tilde C}& =\sum_{k=1}^{K}\sum_{j=1}^{K} \Big|{\tilde G}_{k,j} \Big|^2 \left|{\tilde{\bm u}}_{k,j}^H{\bm \phi}\right|^2
\end{align}
with ${\tilde G}_{k,j} \triangleq G_{R,k}G_{T,j} $ being the combined receive and transmit beamforming gain of the cross link between radar $k$ and radar $j$.
Similar to Section~\ref{Lagrange_single},
 the Lagrange multiplier function for problem (P3) is defined as
\begin{align}
&{\cal L}\left({\bm \theta}, {\bm \lambda}\right)\triangleq  {\bm \theta}^H {\tilde{\bm U}} {\bm \theta} 
+ {\bm \theta}^H {\tilde{\bm v}} +{\tilde{\bm v}}^H{\bm \theta}+ {\tilde C}
+\sum_{n=1}^{N_1} \lambda_n\left( {\theta}_{n}^*{\theta}_n -\beta^2_\text{max} \right)
\notag\\
&= {\bm \theta}^H \left({\tilde{\bm U}}+\text{diag}\left({\bm \lambda}\right) \right) {\bm \theta}
+{\bm \theta}^H {\tilde{\bm v}} +{\tilde{\bm v}}^H{\bm \theta}+ {\tilde C}
-\beta^2_\text{max}{\bm 1}^T{\bm \lambda}\notag
\end{align}
where ${\bm \lambda}\triangleq[\lambda_1, \ldots ,\lambda_{N_1}]^T$ is the corresponding Lagrange multiplier vector with $\lambda_n\ge0, \forall n=1,\ldots,N_1$.
We then take the differentiation of ${\cal L}\left({\bm \theta}, {\bm \lambda}\right)$ with respect to ${\bm \theta}$, yielding
\begin{align}
\frac{\partial{\cal L}\left({\bm \theta}, {\bm \lambda}\right)}{\partial{\bm \theta}}&=\left({\tilde{\bm U}}+\text{diag}\left({\bm \lambda}\right) \right) {\bm \theta} + {\tilde{\bm v}}.
\end{align}
By letting $\frac{\partial{\cal L}\left({\bm \theta}, {\bm \lambda}\right)}{\partial{\bm \theta}}={\bm 0}$, we have
\begin{align}
{\bm \theta}^{\star}=-\left({\tilde{\bm U}}+\text{diag}\left({\bm \lambda}\right) \right)^{-1}
{\tilde{\bm v}} 
\end{align}
which is a function of the Lagrange multiplier vector ${\bm \lambda}$.
Although involving cross links among multiple radars, problem (P3) is still a QCQP with zero-duality gap, implying that the dual
variables $\lambda_n\ge0, \forall n=1,\ldots,N_1$ can be obtained through its dual program.
Accordingly, the dual function is given by 
\begin{align}\label{dual2}
\hspace{-0.2cm}&g\left({\bm \lambda}\right) = \underset{ {\bm \theta} }{\text{inf}} ~{\cal L}\left({\bm \theta}, {\bm \lambda}\right)\notag\\
\hspace{-0.2cm}&\hspace{-0.1cm}=\hspace{-0.15cm}
\left\{
\begin{aligned}
\hspace{-0.2cm}&{\tilde C}\hspace{-0.1cm}-\hspace{-0.1cm}\beta^2_\text{max}{\bm 1}^T{\bm \lambda}\hspace{-0.1cm}-\hspace{-0.1cm}{\tilde{\bm v}}^H \left(\hspace{-0.1cm}{\tilde{\bm U}}\hspace{-0.1cm}+\hspace{-0.1cm}\text{diag}\hspace{-0.1cm}\left({\bm \lambda}\right) \hspace{-0.1cm}\right)^{-1}\hspace{-0.1cm}{\tilde{\bm v}},  {\tilde{\bm U}}\hspace{-0.1cm}+\hspace{-0.1cm}\text{diag}\hspace{-0.1cm}\left({\bm \lambda}\right) \succcurlyeq 0\\
\hspace{-0.2cm}&-\infty, \quad {\rm otherwise}
\end{aligned}
\right.\hspace{-0.1cm}.\hspace{-0.2cm}
\end{align}
Similarly, by exploiting the Schur complement, the dual problem can be expressed as an equivalent semidefinite
optimization problem, i.e., 
\begin{align}
\text{(P3.1):}
& \underset{ q, {\bm \lambda} }{\text{max}}
& &  q\label{obj_P3.1} \\
& \text{s.t.} & & 
\begin{bmatrix}
{\tilde C}-\beta^2_\text{max}{\bm 1}^T{\bm \lambda}-q&{\tilde{\bm v}}^H\\{\tilde{\bm v}}  &{\tilde{\bm U}}+\text{diag}\left({\bm \lambda}\right)
\end{bmatrix}\succcurlyeq 0\label{con1_P3.1}\\
& & & \lambda_n \ge 0, \forall n=1,\ldots,N_1\label{con2_P3.1}
\end{align}
which can be effectively solved via standard SDP or LMI optimization, with the complexity order of ${\cal O}(N_1^{4.5})$ \cite{Luo2010Semidefinite}.

\subsection{MMSE-based Solution} \label{Alignment_multi}
 Upon examining the objective function in \eqref{obj_P1} of problem (P1), we find that the conditions for achieving full electromagnetic stealth against multiple radars are given as follows.
\begin{align}\label{condition2}
\hspace{-0.3cm}{\tilde G}_{k,j}\left({\bm u}_{k,j}^H {\bm \theta} + {\tilde{\bm u}}_{k,j}^H{\bm \phi}\right)=0,\forall k=1,\ldots,K, \forall j=1,\ldots,K\hspace{-0.25cm}
\end{align}
which implies that the IRS's reflection vector ${\bm \theta}$ should be designed to solve the linear equations in \eqref{condition2} to null/offset the signal reflected from the NIRS, thereby achieving electromagnetic stealth.
However, due to the modulus constraints in \eqref{con1_P1}, i.e., $|{\theta}_{n}|\le\beta_\text{max}, \forall n=1,\ldots,N_1$ and the existence of cross links among multiple radars,
the $K^2$ equality conditions in \eqref{condition2} are generally difficult (if not impossible) to attain for all $k$ and $j$ simultaneously.

Inspired by the above, we propose to leverage the 
MMSE-based method to approximately solve the linear equations in \eqref{condition2}.
Specifically, the linear equations in \eqref{condition2} can be rewritten in a compact form as
\begin{align}\label{condition3}
\underbrace{\begin{bmatrix}
{\tilde G}_{1,1}{\bm u}_{1,1}^H \\
\vdots \\
{\tilde G}_{K,K}{\bm u}_{K,K}^H
\end{bmatrix}}_{{\bm D} \in  {\mathbb{C}^{K^2\times N_1 }}}{\bm \theta}
=-\underbrace{\begin{bmatrix}
{\tilde G}_{1,1}{\tilde{\bm u}}_{1,1}^H \\
\vdots \\
{\tilde G}_{K,K}{\tilde{\bm u}}_{K,K}^H
\end{bmatrix}}_{{\bm E} \in  {\mathbb{C}^{K^2\times N_2 }}} {\bm \phi}.
\end{align}
Accordingly, we propose the following MMSE-based solution:
\begin{align}\label{MMSE}
{\bm \theta}_{\rm MMSE}=-\left( {\bm D}^H {\bm D} +\delta {\bm I}_{N_1}\right)^{-1}{\bm D}^H {\bm E} {\bm \phi}
\end{align}
where $\delta$ serves as a regularization parameter, ensuring that the IRS's reflection design ${\bm \theta}$ adheres to the modulus constraints in \eqref{con1_P1}. In particular, $\delta$ can be determined by substituting \eqref{MMSE} into (P1) and
solving the following univariate optimization problem.
\begin{align}
\text{(P4):}
& \underset{ \delta }{\text{min}}
& &   \Big\| -{\bm D}  \left( {\bm D}^H {\bm D} +\delta {\bm I}_{N_1}\right)^{-1}{\bm D}^H {\bm E} {\bm \phi}
 + {\bm E} {\bm \phi} \Big\|^2 \hspace{-0.2cm}\label{obj_P4} \\
& \text{s.t.} & &|{\bm i}_n^T\left( {\bm D}^H {\bm D} +\delta {\bm I}_{N_1}\right)^{-1}{\bm D}^H {\bm E} {\bm \phi}|\le \beta_\text{max}, \forall n \label{con1_P4}
\end{align}
where ${\bm i}_n$ is the $n$-th column of the $N_1\times N_1$ identity matrix ${\bm I}_{N_1}$. It is noted that the optimal $\delta$ in problem (P4) can be readily obtained via one-dimensional search over positive real number field with very low complexity. In addition,  the MMSE-based solution, which involves matrix inversion using methods such as the Gauss-Jordan elimination, leads to a computational complexity of ${\cal O}(N_1^3)$.

\emph{Remark 1:} For the single-radar case, the target only needs to estimate the AoA information from the radar, i.e., $( \vartheta_{R\rightarrow T}, \varphi_{R\rightarrow T})$ to solve problem (P2) optimally. In contrast, for the multi-radar case, the target needs to estimate not only the AoA information from each radar, i.e., $\left\{( \vartheta_{R_k\rightarrow T}, \varphi_{R_k\rightarrow T})\right\}_{k=1}^K$, but also the equivalent transmit/receive beamforming gain at each radar, i.e., $\left\{|G_{T,k}|^2\right\}_{k=1}^K$ and $\left\{|G_{R,k}|^2\right\}_{k=1}^K$, to solve problem (P3) optimally. Moreover, it is worth noting that due to the channel reciprocity of each link in forward and reverse directions, we have $G_{T,k}=G_{R,k}, \forall k=1,\ldots, K$ for the equivalent transmit/receive beamforming gain at each radar. Thus, it is sufficient for the target to estimate the AoA information and transmit beamforming gain from each radar for solving problem (P3) in the multi-radar scenario.

\section{Estimation of AoA and Beamforming Gain at the Target}\label{AoA}
As discussed in the previous section (cf. Remark 1), the target only needs to acquire the AoA information and/or transmit beamforming gain
for the IRS's reflection design to achieve electromagnetic stealth. 
Recall that a cross-shaped array, consisting of $L=L_{x}+ L_{y}-1$ receive sensing devices, is embedded on the TS to enable its sensing mode for radar reconnaissance.
Specifically, we use the subscript ``$S$" to indicate the CSSA and 
let ${\bm a}_{S}(\vartheta_{R_k\rightarrow T}, \varphi_{R_k\rightarrow T})$ denote its array response vector, which can be decomposed into two 1D steering vectors along the $x$- and $y$-axes, respectively, as given by
\begin{align}
\hspace{-0.25cm}{\bm a}_{S,x}(\vartheta_{R_k\rightarrow T},\hspace{-0.05cm} \varphi_{R_k\rightarrow T})&\hspace{-0.1cm}=\hspace{-0.1cm} {\bm e}\hspace{-0.1cm}\left(\hspace{-0.1cm} \frac{2\Delta_e}{\lambda} \hspace{-0.05cm}\cos (\varphi_{R_k\rightarrow T})\hspace{-0.05cm}\cos (\vartheta_{R_k\rightarrow T})  ,\hspace{-0.05cm} L_{x} \hspace{-0.1cm}\right) \hspace{-0.1cm} e^{\varphi_x}\notag\\
\hspace{-0.25cm}{\bm a}_{S,y}(\vartheta_{R_k\rightarrow T},\hspace{-0.05cm} \varphi_{R_k\rightarrow T})&\hspace{-0.1cm}=\hspace{-0.1cm}{\bm e}\hspace{-0.1cm}\left(\hspace{-0.1cm}\frac{2\Delta_e}{\lambda} \hspace{-0.05cm}\cos(\varphi_{R_k\rightarrow T})\hspace{-0.05cm}\sin(\vartheta_{R_k\rightarrow T}) ,\hspace{-0.05cm} L_{y} \hspace{-0.1cm}\right)\hspace{-0.1cm} e^{ \varphi_y } \notag
\end{align}
where $\varphi_x\triangleq\frac{-j \pi (L_{x}-1)\Delta_e}{\lambda} \cos (\varphi_{R_k\rightarrow T})\cos (\vartheta_{R_k\rightarrow T})$ and
$\varphi_y\triangleq\frac{-j \pi (L_{y}-1)\Delta_e}{\lambda} \cos(\varphi_{R_k\rightarrow T})\sin (\vartheta_{R_k\rightarrow T})$ are the phase-shift offsets that we introduce to ensure the same response at the (central) common sensing device along the $x$- and $y$-axes. Accordingly, the channel between the CSSA and radar~$k$, denoted by ${\bm H}_{R_k\rightarrow S}\in {\mathbb{C}^{L\times M }}$, can be similarly
modeled as the outer product of array responses at their two sides as \eqref{Far_LoS} by setting $X\in\{S\}$ and $Y \in \{R_k\}_{k=1}^K$.
Moreover, as the CSSA is embedded on the TS, it shares the same AoA pair of $(\vartheta_{R_k\rightarrow T}, \varphi_{R_k\rightarrow T})$ and the same path gain of $\rho_{{R_k}-T}$.

Based on the above and with the radar pulse waveform given in \eqref{waveform}, the signals received at the CSSA 
at time $t$ can be expressed as
\begin{align}\label{cross-shaped}
&\hspace{-0.35cm}{\bm z}(t)=
\sum_{k=1}^{K} {\bm H}_{R_k\hspace{-0.05cm}\rightarrow\hspace{-0.05cm} S} {\bm w}_k{x}_k(t)+{\bm n}_{S}(t)\notag\\
&\hspace{-0.35cm}=\hspace{-0.15cm}\sum_{k=1}^{K} \hspace{-0.1cm}{\bm a}_{S}\hspace{-0.05cm}(\hspace{-0.05cm} \vartheta_{R_k\hspace{-0.05cm}\rightarrow\hspace{-0.05cm} T},\hspace{-0.05cm} \varphi_{R_k\hspace{-0.05cm}\rightarrow\hspace{-0.05cm} T}\hspace{-0.05cm})\hspace{-0.1cm}  \underbrace{\rho_{{R_k}-T} {\bm a}^T_{R_k}\hspace{-0.05cm}(\hspace{-0.05cm} \vartheta_{T\hspace{-0.05cm}\rightarrow\hspace{-0.05cm} R_k}, \hspace{-0.05cm}\varphi_{T\hspace{-0.05cm}\rightarrow\hspace{-0.05cm} R_k}\hspace{-0.05cm})  {\bm w}_k}_{G_{T,k}:\rm transmit~beamforming~gain} \hspace{-0.1cm} {x}_k(\hspace{-0.05cm}t\hspace{-0.05cm})\hspace{-0.1cm}+\hspace{-0.1cm}{\bm n}_{S}(\hspace{-0.05cm}t\hspace{-0.05cm})\notag\\
&\hspace{-0.35cm}=\hspace{-0.1cm}\underbrace{\begin{bmatrix}
	{\bm a}_{S}( \vartheta_{R_1\hspace{-0.05cm}\rightarrow\hspace{-0.05cm} T}, \varphi_{R_1\hspace{-0.05cm}\rightarrow\hspace{-0.05cm} T}),\cdots,{\bm a}_{S}( \vartheta_{R_K\hspace{-0.05cm}\rightarrow\hspace{-0.05cm} T}, \varphi_{R_K\hspace{-0.05cm}\rightarrow\hspace{-0.05cm} T})
	\end{bmatrix}}_{{\bm A}_{S}}{\bm s}(t)+{\bm n}_{S}(t)
\end{align}
where ${\bm n}_{S}(t)\sim {\mathcal N_c }({\bm 0}, \sigma_{S}^2{\bm I}_L)$ is the AWGN vector at the CSSA
with $\sigma_{S}^2$ being the noise power,
${\bm A}_{S}$ denotes the array response matrix at the CSSA,
and ${\bm s}(t)\triangleq\left[ {s}_1(t), \ldots ,{s}_K(t)\right]^T$ denotes the transmit beamforming signal vector from $K$ radars, with ${s}_k(t)=G_{T,k} {x}_k(t), \forall k=1,\ldots,K$. 

\subsection{Single-Radar Case}
For the single-radar setup, i.e., $K = 1$, the signal vector received at the CSSA reduces to
\begin{align}\label{cross-shaped_xy1}
{\bm z}(t)=
 G_{T} {x}(t) {\bm a}_{S}( \vartheta_{R\rightarrow T}, \varphi_{R\rightarrow T}) +{\bm n}_{S}(t).
\end{align}
Based on \eqref{cross-shaped_xy1}, existing AoA estimation
algorithms such as multiple signal classification (MUSIC) \cite{Swindlehurst1992MUSIC} can be applied to estimate the AoA pair $( \vartheta_{R\rightarrow T}, \varphi_{R\rightarrow T})$.
Then, with the estimated AoA information, the IRS's reflection can be practically designed according to Section~\ref{Single-Radar} to achieve electromagnetic stealth against the single radar.

\subsection{Multi-Radar Case}
Based on \eqref{cross-shaped} for the multi-radar case, existing AoA estimation algorithms such as MUSIC can also be applied to estimate multiple AoA pairs from different radars, i.e., $\left\{( \vartheta_{R_k\rightarrow T}, \varphi_{R_k\rightarrow T})\right\}_{k=1}^K$.
Note that in addition to the AoA information, the target also needs to estimate the equivalent transmit/receive beamforming gain at each radar, i.e., $\left\{|G_{T,k}|^2\right\}_{k=1}^K$ and $\left\{|G_{R,k}|^2\right\}_{k=1}^K$ for the multi-radar case, to practically solve problem (P3). Specifically, based on \eqref{cross-shaped} and with the estimated AoA pairs to obtain ${\bm A}_{S}$, the least-squares (LS) estimate of ${\bm s}(t)$ is given by
\begin{align}
\hspace{-0.15cm}{\hat{\bm s}}(t)\hspace{-0.1cm}=\hspace{-0.1cm}\left({\bm A}_{S}^H{\bm A}_{S} \right)^{-1}\hspace{-0.1cm} {\bm A}_{S}^H {\bm z}(t)\hspace{-0.1cm}=\hspace{-0.1cm}{\bm s} (t)
\hspace{-0.1cm}+\hspace{-0.1cm}\underbrace{\left({\bm A}_{S}^H{\bm A}_{S} \right)^{-1}\hspace{-0.1cm} {\bm A}_{S}^H {\bm n}_{S}(t)}_{{\bar{\bm n}}_{S}(t)}\hspace{-0.2cm}
\end{align}
where ${\hat{\bm s}}(t)=\left[ {\hat s}_1(t), \ldots ,{\hat s}_K(t)\right]^T$ and ${\bar{\bm n}}_{S}(t)=\left[ {\bar n}_{S,1}(t), \ldots ,{\bar n}_{S,K}(t)\right]^T$ is the effective noise vector for the LS estimation. Accordingly, the estimate of transmit beamforming gain $|G_{T,k}|^2$ from each radar over one PRI is given by
\begin{align}\label{LS_est}
&|{\hat G}_{T,k}|^2=\frac{1}{T_p}\int_{0}^{T_p} |{\hat s}_k(t)|^2 {\rm d}t 
=\frac{1}{T_p}\int_{0}^{T_p} |G_{T,k} {x}_k(t) + {\bar n}_{S,k}(t) |^2 {\rm d}t \notag\\
&=\frac{1}{T_p}\int_{0}^{T_p} |G_{T,k}|^2 |x_k(t)|^2  + 
2\Re\{G_{T,k}x_k(t) {\bar n}^*_{S,k}(t) \}
+|{\bar n}_{S,k}(t)|^2 {\rm d}t\notag\\
&=P|G_{T,k}|^2+ \frac{1}{T_p}\int_{0}^{T_p} |{\bar n}_{S,k}(t)|^2  {\rm d}t.
\end{align}
From \eqref{LS_est}, one can observe that the transmit beamforming gain $|G_{T,k}|^2$ is estimated with a scaling ambiguity of $P$, which, however, does not affect the solution to problem (P3) in the multi-radar case.
On the other hand, the estimate of the receive beamforming gain $|G_{R,k}|^2$ can be derived from the estimated transmit beamforming gain by leveraging the channel reciprocity of each link in both forward and reverse directions, i.e., $|G_{R,k}|^2=|G_{T,k}|^2, \forall k=1,\ldots,K$.
Finally, with the estimated AoA and transmit/receive beamforming gains in \eqref{LS_est}, the IRS's reflection can be designed according to Section~\ref{Multi-Radar} to achieve electromagnetic stealth against multiple radars.

\section{Simulation Results}\label{Sim}
\begin{figure}[!t]
	\centering
	\includegraphics[width=3.5in]{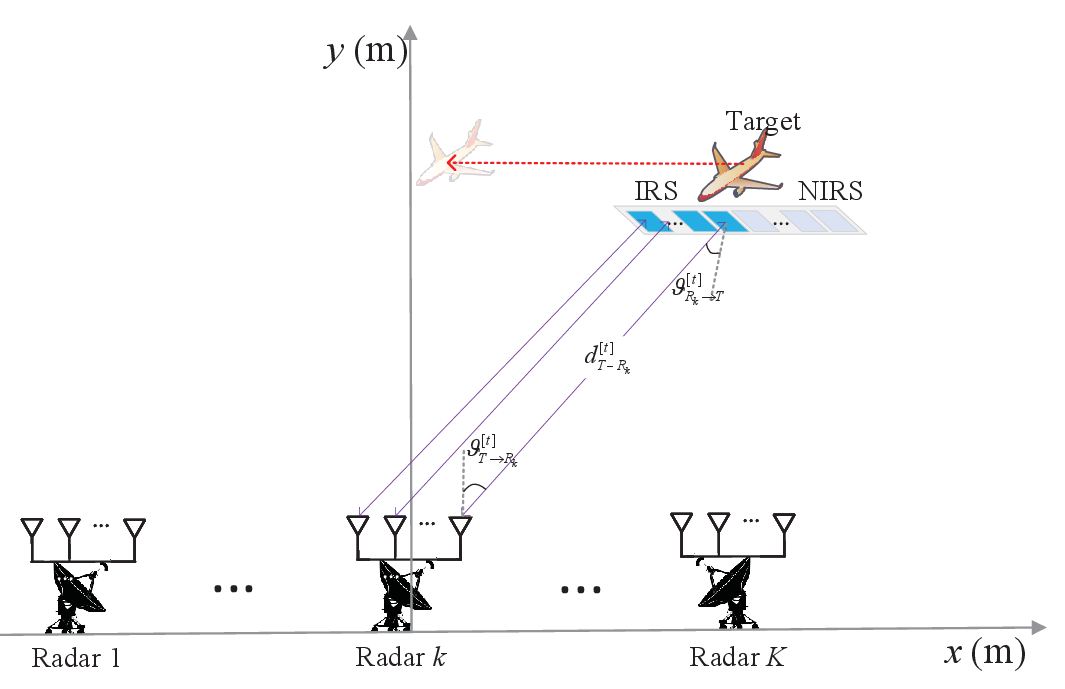}
	\setlength{\abovecaptionskip}{-3pt}
	\caption{An illustration for the geometric relationship between the target and
		radars (2D view).}
	\label{ESS_sim}
\end{figure}
In this section, we present simulation results to evaluate the performance of the proposed IRS-aided electromagnetic stealth system against radar detection, as well as the proposed algorithms for the IRS's reflection design to achieve electromagnetic stealth. As depicted in Fig.~\ref{ESS_sim},
we consider that the moving target (e.g., an aircraft) and the radar(s) are situated within the same 2D plane. In particular, under the considered setup, we have $\varphi_{R_k\rightarrow T}^{[t]}=\varphi_{T\rightarrow R_k}^{[t]}=0$ and thus we only need to focus on the real-time variations of AoAs/AoDs $\left\{\vartheta_{T\rightarrow R_k}^{[t]}, \vartheta_{R_k\rightarrow T}^{[t]}\right\} \in (-\pi/2,\pi/2)$ in our simulations.
Each mono-static radar is equipped with a UPA consisting of $M= 8\times 8=64$ transmit/receive antennas; the IRS and NIRS are also assumed to be UPAs with $N_1= N_{1,x} \times 2$ and $N_2=100 \times 2=200$ passive elements, respectively, with $N_1$ or $N_{1,x}$ to be specified in the simulations; and the CSSA is composed of $L=9$ sensing devices. The absorbing efficiency of the NIRS elements is set as $\zeta_n=0.8, \forall n=1,\ldots,N_2$ and the maximum reflection amplitude of each IRS element is set as $\beta_\text{max}= 1$. Moreover, the (non-tunable) phase shifts of the NIRS, i.e., $\{\phi_{n}\}_{n=1}^{N_2}$ are randomly generated following a uniform distribution within $[0,2\pi)$.

We assume that the single/multi-radar system operates at the super high frequency (SHF) of 150 gigahertz
(GHz) with the wavelength of $\lambda=0.05$ m. Moreover, the antenna spacing at each radar and
the element spacing at the IRS/NIRS/TS are set as $\Delta_a=\lambda/2=0.025$ m and $\Delta_e=\lambda/4=0.0125$ m, respectively.
 The radar pulse waveform is given by $x(t)=\sqrt{P} e^{(j \pi B t^2/t_p)}$ with the signal bandwidth of $B = 100$ MHz.
In addition, the PRI is set as $T_p=100~\mu$s with the pulse duration of $t_p=30~\mu$s.
Unless otherwise specified, the reference path gain at the distance of $1$ m is set as $\alpha=-30$ dB for each individual link; the shortest radar-target distance is set as $100$ m; and the transmit power of each radar is set as $P=15$ dBm.

For the purpose of exposition, we first assume that the perfect information in terms of AoA and transmit/receive beamforming gain is available at the target; while the effect of imperfect information (estimated at the CSSA of the target) on the radar performance will be investigated subsequently.
Moreover, for the reflection design in the IRS-aided electromagnetic stealth system, we consider two benchmark schemes as follows.
 \begin{itemize}
 	\item {\bf Random phase-shift design}: In this scheme, the phase shifts
 	in ${\bm \theta}^{[t]}$ are randomly generated following a uniform distribution within $[0,2\pi)$ at the IRS.
 	\item {\bf DFT-based codebook search}: In this scheme, the IRS's reflection vector ${\bm \theta}^{[t]}$ is searched over a discrete Fourier transform (DFT)-based codebook (denoted by ${\cal D}$) to minimize the (sum) received signal power of the radar(s) in \eqref{obj_P2} and \eqref{obj_P1}, i.e., $\underset{ {\bm \theta} \in {\cal D} }{\text{min}}  \Big| {\bm u}^H {\bm \theta} + C \Big|^2$ for the single-radar case and
 	$\underset{ {\bm \theta} \in {\cal D} }{\text{min}}
 	\sum_{k=1}^{K}\sum_{j=1}^{K} \Big|G_{R,k} \Big|^2 \Big|G_{T,j} \Big|^2 \Big| {\bm u}_{k,j}^H {\bm \theta} + {\tilde{\bm u}}_{k,j}^H{\bm \phi} \Big|^2$ for the multi-radar case, respectively.
 \end{itemize}

\subsection{Single-Radar Case}
\begin{figure}[!t]
	\centering
	\includegraphics[width=3.5in]{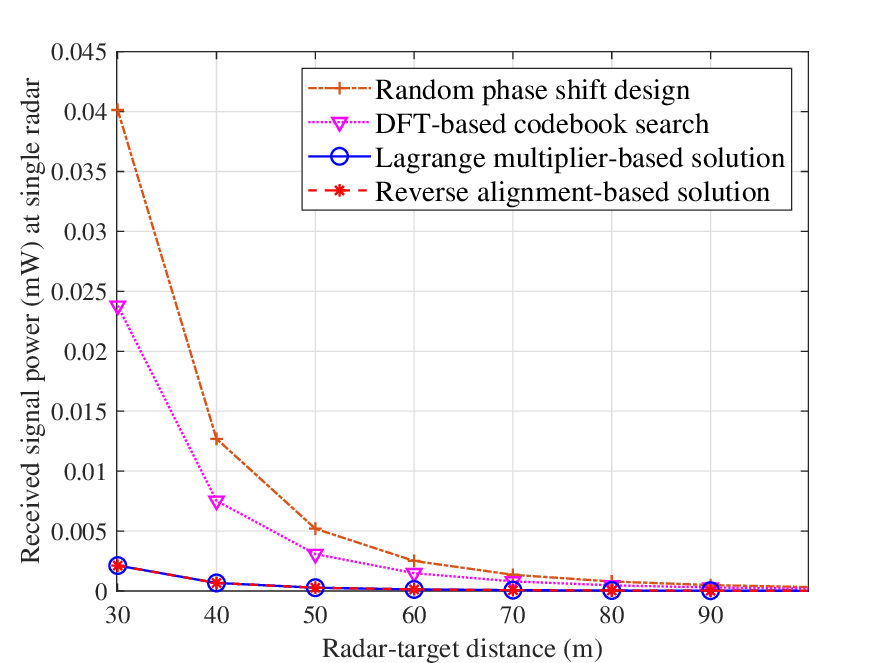}
	\setlength{\abovecaptionskip}{-5pt}
	\caption{Single radar's received signal power versus radar-target distance, with $N_1=4\times2=8$.}
	\label{gain_vs_distance_sRadar3}
\end{figure}
First, we consider the single-radar case (i.e., $K=1$) with $N_1=4\times2=8$ tunable passive reflecting elements at the IRS. In Fig.~\ref{gain_vs_distance_sRadar3}, we compare the received signal power of the single radar against the radar-target distance for different reflection designs in the IRS-aided electromagnetic stealth system.
It is observed that, irrespective of the radar-target distance, the proposed Lagrange multiplier- and reverse alignment-based solutions consistently achieve the same optimal performance. Both result in a significantly lower received signal power at the single radar compared to the random phase-shift design and DFT-based codebook search. This validates the effectiveness of the proposed solutions for the IRS's reflection design in reducing or eliminating the reflected signal power, thereby preventing detection by the single adversary radar. On the other hand, by searching over a given codebook ${\cal D}$, the DFT-based codebook search can reduce almost half of the received signal power compared to the random phase-shift design.

\begin{figure}[!t]
	\centering
	\includegraphics[width=3.5in]{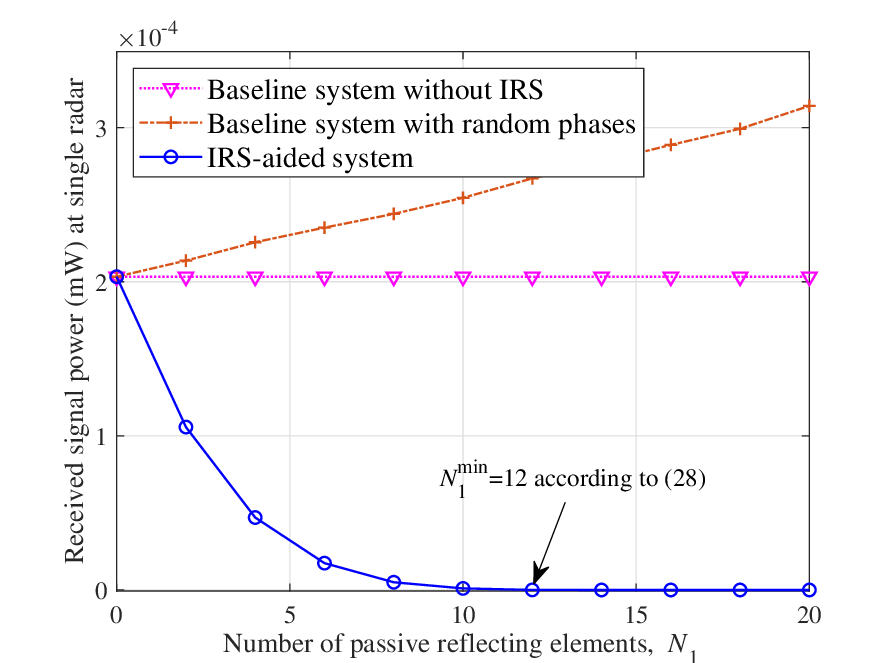}
	\setlength{\abovecaptionskip}{-5pt}
	\caption{Single radar's received signal power versus the number of tunable passive reflecting elements $N_1$ at the target.}
	\label{gain_vs_element_sRadar2}
\end{figure}
In Fig.~\ref{gain_vs_element_sRadar2}, we show the single radar's received signal power versus the number of tunable passive reflecting elements $N_1$ at the target equipped with different electromagnetic stealth systems. For comparison, we consider two baseline systems: 1) The baseline system without IRS, equivalent to setting ${\bm \theta}^{[t]}={\bm 0}$; and 2) The baseline system with random phase-shift design at the IRS. Several key observations can be made from Fig.~\ref{gain_vs_element_sRadar2}.
First, as the number of tunable passive reflecting elements $N_1$ increases in the IRS-aided electromagnetic stealth system, the single radar's received signal power decreases until reaching the minimum value of $0$, which is due to the increased capability of signal nulling/cancellation at the IRS.
Second, the IRS-aided electromagnetic stealth system achieves full electromagnetic stealth (i.e., the received signal power ${\bar P}=0$ at the radar) when $N_1\ge 12$. This corroborates the accuracy of the analytical results for the minimum number of IRS elements in \eqref{N_1min} as $N_1^{\min}=12$, which is significantly smaller compared to the number of non-tunable passive reflecting elements $N_2=200$ in our setting.
Finally, in contrast to the IRS-aided electromagnetic stealth system, the two baseline systems do not decrease the radar's received signal power by increasing $N_1$.
This is expected since we have ${\bm u}^H {\bm \theta}=0$ in the baseline system without IRS; while the baseline system with random phase-shift design at the IRS will increase the radar's received signal power,
due to more reflection power radiated from the target without properly designing the IRS for signal cancellation.

\begin{figure}[!t]
	\centering
	\includegraphics[width=3.5in]{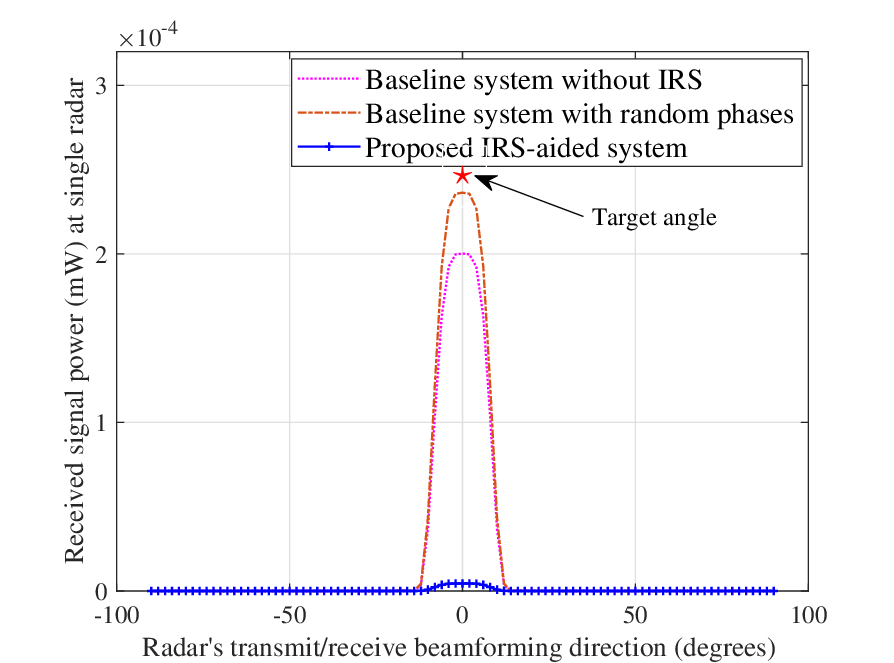}
	\setlength{\abovecaptionskip}{-5pt}
	\caption{Single radar's received signal power versus its transmit/receive beamforming direction, with $N_1=4\times2=8$.}
	\label{gain_vs_angle_sRadar2}
\end{figure}
In Fig.~\ref{gain_vs_angle_sRadar2}, we plot the single radar's received signal power versus its transmit/receive beamforming direction in different electromagnetic stealth systems, with $N_1=4\times2=8$ tunable passive reflecting elements. In the simulation setting as shown in Fig.~\ref{ESS_sim}, the target is positioned directly above the radar with $\vartheta_{T\rightarrow R}^{[t]}=0^\circ$. It is observed that at the target angle of $\vartheta_{T\rightarrow R}^{[t]}=0^\circ$, the received signal power at the radar can be significantly reduced by deploying the IRS-aided electromagnetic stealth system on the target, as compared to the two baseline systems without IRS and with random phase-shift design at the IRS. This thus validates the effectiveness of the proposed IRS-aided system in achieving electromagnetic stealth for the target, by meticulously designing the IRS reflection to null/mitigate the reflected signal to the single radar.

\begin{figure}[!t]
	\centering
	\includegraphics[width=3.5in]{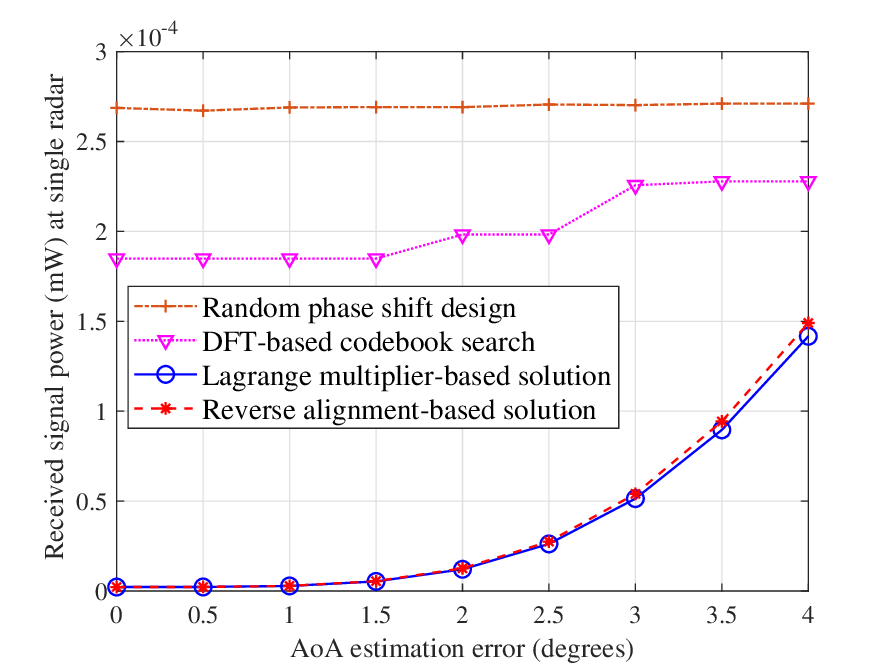}
	\setlength{\abovecaptionskip}{-5pt}
	\caption{Effect of imperfect AoA information at the target to single radar's received signal power, with $N_1=4\times2=8$.}
	\label{gain_vs_angle_err_sRadar}
\end{figure}
In Fig.~\ref{gain_vs_angle_err_sRadar}, we examine the impact of imperfect AoA information at the target on the received signal power at the single radar, under different reflection designs in the IRS-aided electromagnetic stealth system.
One can observe that as the AoA estimation error at the CSSA increases, all schemes (except the random phase-shift design) result in an increase in the radar's received signal power. This increase is attributed to the imperfect signal nulling/cancellation at the IRS, leading to more signal power being radiated from the target. This suggests that in practice, the AoA estimation error will diminish the signal nulling/cancellation capability at the IRS.
Nevertheless, when the AoA estimation error is below $2^\circ$ (which approximately corresponds to a localization error of less than $3.5$ m given a radar-target distance of $100$ m), the proposed Lagrange multiplier- and reverse alignment-based solutions still maintain a remarkably low level of received signal power at the single radar. Specifically, for the AoA estimation error at $2^\circ$, the received signal power level is merely about $6.3\%$ and $4.7\%$ of that of the DFT-based codebook search and random phase-shift design, respectively.

\subsection{Multi-Radar Case}

\begin{figure}[!t]
	\centering
	\includegraphics[width=3.5in]{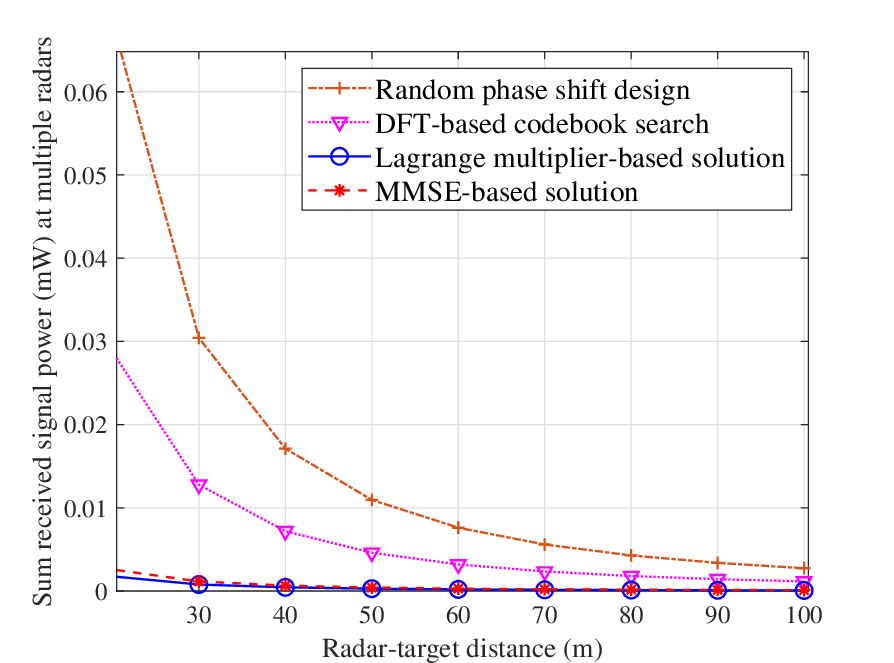}
	\setlength{\abovecaptionskip}{-5pt}
	\caption{Multi-radars' sum received signal power versus radar-target distance, with $N_1=25\times2=50$.}
	\label{gain_vs_distance_mRadar}
\end{figure}
Next, we delve into the multi-radar case involving three radars ($K=3$) with their locations illustrated in Fig.~\ref{ESS_sim}, where the target is equipped with the IRS of $N_1=25\times2=50$ tunable passive reflecting elements.
In Fig.~\ref{gain_vs_distance_mRadar}, we compare the multi-radars' sum received signal power versus the shortest radar-target distance, under different reflection designs in the IRS-aided electromagnetic stealth system.
Our observations reveal that the proposed MMSE-based solution, despite its lower complexity, closely approximates the optimal Lagrange multiplier-based solution. Moreover, when compared to the random phase-shift design and DFT-based codebook search, both proposed solutions result in a significantly lower signal power received at multiple radars. 
This validates their effectiveness in the IRS's reflection design for reducing or even eliminating the reflected signal power to multiple radars simultaneously. On the other hand, by searching over a predefined codebook ${\cal D}$, the DFT-based codebook search can reduce more than half of the received signal power compared to the random phase-shift design.

\begin{figure}[!t]
	\centering
	\includegraphics[width=3.5in]{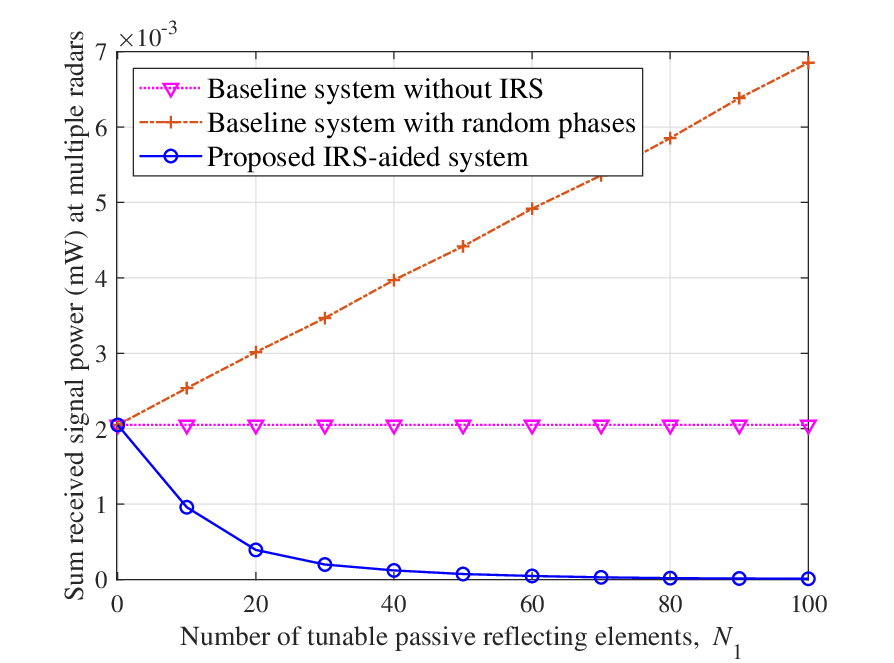}
	\setlength{\abovecaptionskip}{-5pt}
	\caption{Multi-radars' sum received signal power versus the number of passive reflecting elements $N_1$ at the target.}
	\label{gain_vs_element_mRadar2}
\end{figure}
In Fig.~\ref{gain_vs_element_mRadar2}, we show the multi-radars' sum received signal power versus the number of tunable passive reflecting elements $N_1$ at the target. Several interesting observations can be made as follows.
First, an increase in the number of tunable passive reflecting elements $N_1$ in the IRS-aided electromagnetic stealth system leads to a decrease in the sum received signal power at multiple radars. This is accomplished by harnessing the enhanced signal cancellation capability at the IRS to null or mitigate the reflected signal to multiple radars. 
Second, even in the more challenging multi-radar case that involves cross links among multiple radars, the proposed IRS-aided electromagnetic stealth system can still attain full electromagnetic stealth (i.e., the received signal power ${\bar P}_k \approxeq0$ at each radar) when $N_1\ge 90$. In contrast to the single-radar case, which necessitates only one equality condition to attain full electromagnetic stealth, the multi-radar case requires a considerably larger minimum number of IRS elements to meet  the $K\times K=3\times3= 9$ equality conditions as per \eqref{condition2}.
Lastly, unlike the IRS-aided electromagnetic stealth system, the baseline system without IRS remains unchanged with respect to $N_1$ as ${\bm u}^H {\bm \theta}=0$. However, due to the increased signal power reflected from the IRS without proper signal cancellation, the baseline system with random phase-shift design at the IRS experiences a significant increase in the sum received signal power at multiple radars as $N_1$ increases.

\begin{figure}[!t]
	\centering
	\includegraphics[width=3.5in]{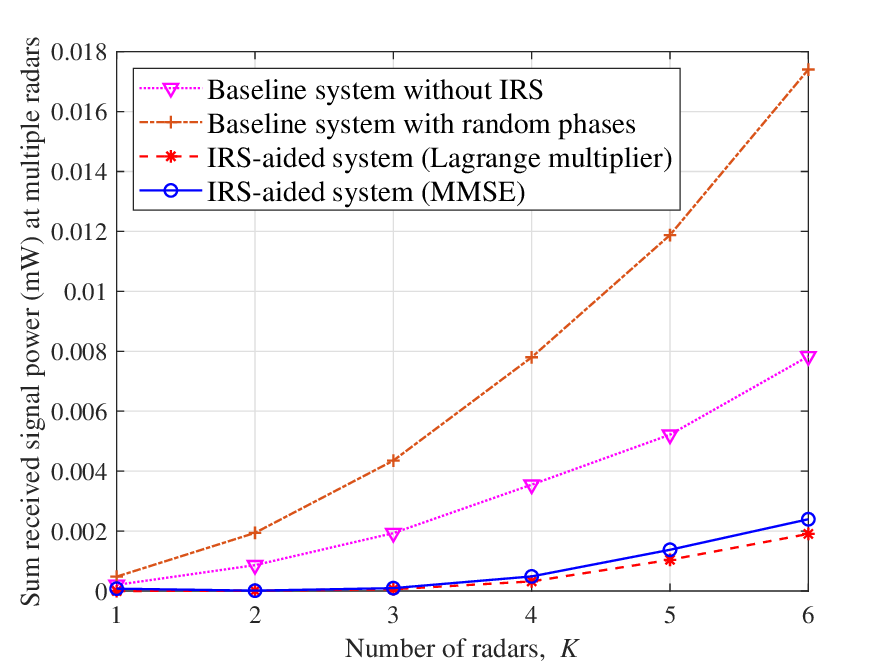}
	\setlength{\abovecaptionskip}{-5pt}
	\caption{Sum received signal power versus the number of radars, with $N_1=25\times2=50$.}
	\label{gain_vs_NUMradar_mRadar}
\end{figure}
In Fig.~\ref{gain_vs_NUMradar_mRadar}, we show the multi-radars' sum received signal power versus the number of radars,
for the comparison among different electromagnetic stealth systems.
It is observed that all the systems experience an increase in the sum received signal power at multiple radars as the number of radars increases. This is expected since a larger number of distributed radars inherently implies a broader collective reception capacity from different directions, thereby leading to an increase in the sum received signal power. 
As compared to the two baseline systems without IRS and with random phase-shift design, the proposed IRS-aided electromagnetic stealth system exhibits a significantly slower rate of increase in the sum received signal power as $K$ increases. This can be attributed to the effectiveness of the proposed IRS's reflection designs (based on both Lagrange multiplier and MMSE) in nulling or mitigating the reflected signal over multiple directions to several radars simultaneously.

\begin{figure}[!t]
	\centering
	\includegraphics[width=3.5in]{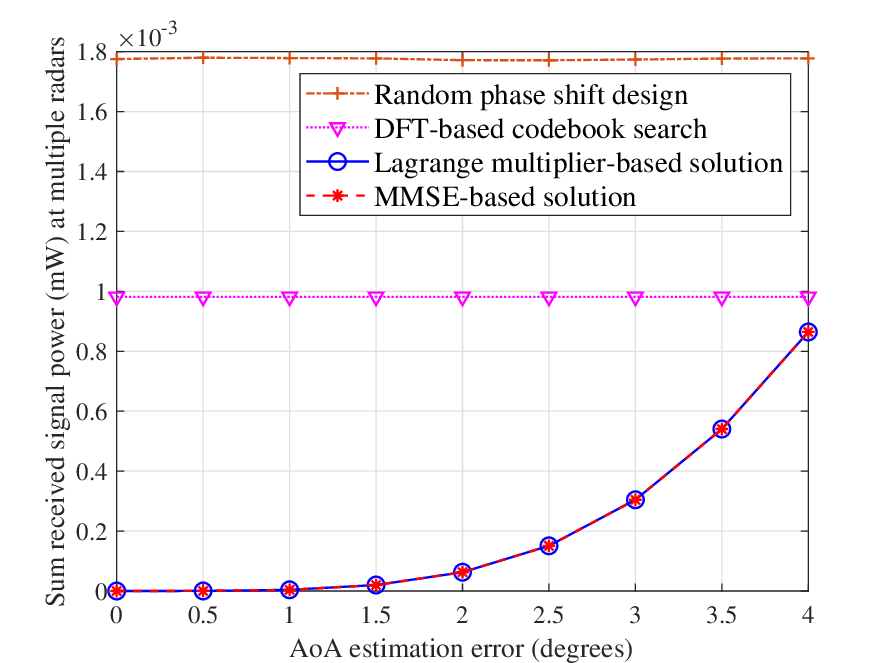}
	\setlength{\abovecaptionskip}{-5pt}
	\caption{Effect of imperfect AoA information at the target to multi-radars' sum received signal power, with $N_1=25\times2=50$.}
	\label{gain_vs_angle_err_mRadar}
\end{figure}
In Fig.~\ref{gain_vs_angle_err_mRadar}, we study the effect of imperfect AoA information at the target on the multi-radars' sum received signal power, under different reflection designs in the IRS-aided electromagnetic stealth system.
It is observed that for the proposed Lagrange multiplier- and MMSE-based solutions, an increase in the AoA estimation error at the CSSA leads to an increase in the sum received signal power at multiple radars. This is due to the AoA estimation error causing imperfect signal nulling/cancellation at the IRS, which in turn results in an increased signal power reflected from the target to multiple radars. 
Fortunately, when the AoA estimation error is no more than $2^\circ$, the proposed Lagrange multiplier- and MMSE-based solutions continue to maintain a very low received signal power level at multiple radars. Specifically, their corresponding received signal power is only about $6.4\%$ and $3.5\%$ of that of the DFT-based codebook search and random phase-shift design, respectively.

\section{Conclusions}\label{conlusion}
In this paper, we studied a new IRS-aided adaptive electromagnetic stealth system for a target to evade radar detection. We formulated an optimization problem for designing the IRS's reflection at the target with the goal of minimizing the sum of received signal power over all adversary radars. Utilizing the Lagrange multiplier method, we derived a semi-closed-form optimal solution for the single-radar setup and extended it to the multi-radar case. Furthermore, we proposed low-complexity closed-form solutions based on reverse alignment/cancellation and MMSE criteria, tailored for the single-radar and multi-radar cases, respectively. 
In particular, we also analyzed the minimum number of IRS elements required to achieve full electromagnetic stealth in the single-radar case. 
To acquire AoA and/or path gain information, we developed practical low-complexity estimation schemes with the aid of receive sensing devices at the target. Simulation results validated the performance advantages of our proposed IRS-aided electromagnetic stealth system using the proposed IRS reflection designs, and demonstrated its performance superiority compared to various baseline systems.

\ifCLASSOPTIONcaptionsoff
  \newpage
\fi

\bibliographystyle{IEEEtran}
\bibliography{IRS_Stealth}

\begin{thebibliography}{10}
\providecommand{\url}[1]{#1}
\csname url@samestyle\endcsname
\providecommand{\newblock}{\relax}
\providecommand{\bibinfo}[2]{#2}
\providecommand{\BIBentrySTDinterwordspacing}{\spaceskip=0pt\relax}
\providecommand{\BIBentryALTinterwordstretchfactor}{4}
\providecommand{\BIBentryALTinterwordspacing}{\spaceskip=\fontdimen2\font plus
\BIBentryALTinterwordstretchfactor\fontdimen3\font minus
  \fontdimen4\font\relax}
\providecommand{\BIBforeignlanguage}[2]{{%
\expandafter\ifx\csname l@#1\endcsname\relax
\typeout{** WARNING: IEEEtran.bst: No hyphenation pattern has been}%
\typeout{** loaded for the language `#1'. Using the pattern for}%
\typeout{** the default language instead.}%
\else
\language=\csname l@#1\endcsname
\fi
#2}}
\providecommand{\BIBdecl}{\relax}
\BIBdecl

\bibitem{westra2009radar}
A.~G. Westra, ``Radar versus stealth: Passive radar and the future of {US}
  military power,'' \emph{Joint Force Quarterly: JFQ}, no.~55, pp. 136--143,
  Fourth quarter 2009.

\bibitem{zikidis2014low}
K.~Zikidis, A.~Skondras, and C.~Tokas, ``Low observable principles, stealth
  aircraft and anti-stealth technologies,'' \emph{J. Comput. Model.}, vol.~4,
  no.~1, pp. 129--165, 2014.

\bibitem{rao2002integrated}
G.~Rao and S.~Mahulikar, ``Integrated review of stealth technology and its role
  in airpower,'' \emph{Aeronaut. J.}, vol. 106, no. 1066, pp. 629--642, Dec.
  2002.

\bibitem{yuan2011properties}
C.-X. Yuan, Z.-X. Zhou, J.~W. Zhang, X.-L. Xiang, Y.~Feng, and H.-G. Sun,
  ``Properties of propagation of electromagnetic wave in a multilayer
  radar-absorbing structure with plasma- and radar-absorbing material,''
  \emph{IEEE Trans. Plasma Sci.}, vol.~39, no.~9, pp. 1768--1775, Sept. 2011.

\bibitem{vinoy1996radar}
K.~J. Vinoy and R.~M. Jha, \emph{Radar Absorbing Materials: From Theory to
  Design and Characterization}.\hskip 1em plus 0.5em minus 0.4em\relax
  Springer, 1996.

\bibitem{feng2006electromagnetic}
Y.-B. Feng, T.~Qiu, C.-Y. Shen, and X.-Y. Li, ``Electromagnetic and absorption
  properties of carbonyl iron{/}rubber radar absorbing materials,'' \emph{IEEE
  Trans. Magn.}, vol.~42, no.~3, pp. 363--368, Mar. 2006.

\bibitem{ahmad2019stealth}
H.~Ahmad \emph{et~al.}, ``Stealth technology: Methods and composite materials:
  A review,'' \emph{Polym. Compos.}, vol.~40, no.~12, pp. 4457--4472, Jun.
  2019.

\bibitem{micheli2014synthesis}
D.~Micheli, A.~Vricella, R.~Pastore, and M.~Marchetti, ``Synthesis and
  electromagnetic characterization of frequency selective radar absorbing
  materials using carbon nanopowders,'' \emph{Carbon}, vol.~77, pp. 756--774,
  Jun. 2014.

\bibitem{bai2015reflections}
B.~Bai, X.~Li, J.~Xu, and Y.~Liu, ``Reflections of electromagnetic waves
  obliquely incident on a multilayer stealth structure with plasma and radar
  absorbing material,'' \emph{IEEE Trans. Plasma Sci.}, vol.~43, no.~8, pp.
  2588--2597, Aug. 2015.

\bibitem{wang2022structural}
W.~Wang, D.~Liu, H.~Cheng, T.~Cao, Y.~Li, Y.~Deng, and W.~Xie, ``Structural
  design and broadband radar absorbing performance of multi-layer patch using
  carbon black,'' \emph{Adv. Compos. and Hybrid Mater.}, vol.~5, pp. 1--9, Dec.
  2022.

\bibitem{liu2016novel}
L.~Liu, S.~Liu, H.~Zhang, X.~Kong, and L.~Wang, ``A novel broadband
  metamaterial absorber based on cross dumbbell-shaped structure,'' in
  \emph{Proc. IEEE Int. Wkshps. Electromagnetics (IWEM)}, Nanjing, China, Jul.
  2016, pp. 1--2.

\bibitem{alaee2017theory}
R.~Alaee, M.~Albooyeh, and C.~Rockstuhl, ``Theory of metasurface based perfect
  absorbers,'' \emph{J. Phys. D: Appl. Phys.}, vol.~50, no.~50, pp.
  503\,002--503\,017, Nov. 2017.

\bibitem{cui2014coding}
T.~J. Cui, M.~Q. Qi, X.~Wan, J.~Zhao, and Q.~Cheng, ``Coding metamaterials,
  digital metamaterials and programmable metamaterials,'' \emph{L. Sci. \&
  Appl.}, vol.~3, no.~10, pp. e218--e226, Oct. 2014.

\bibitem{liaskos2018new}
C.~Liaskos, S.~Nie, A.~Tsioliaridou, A.~Pitsillides, S.~Ioannidis, and
  I.~Akyildiz, ``A new wireless communication paradigm through
  software-controlled metasurfaces,'' \emph{IEEE Commun. Mag.}, vol.~56, no.~9,
  pp. 162--169, Sept. 2018.

\bibitem{liu2018programmable}
F.~Liu \emph{et~al.}, ``Programmable metasurfaces: State of the art and
  prospects,'' in \emph{Proc. IEEE Int. Symp. Circuits Syst. (ISCAS)},
  Florence, Italy, May 2018, pp. 1--5.

\bibitem{wu2021intelligent}
Q.~Wu, S.~Zhang, B.~Zheng, C.~You, and R.~Zhang, ``Intelligent reflecting
  surface aided wireless communications: A tutorial,'' \emph{IEEE Trans.
  Commun.}, vol.~69, no.~5, pp. 3313--3351, May 2021.

\bibitem{qingqing2019towards}
Q.~Wu and R.~Zhang, ``Towards smart and reconfigurable environment: Intelligent
  reflecting surface aided wireless network,'' \emph{IEEE Commun. Mag.},
  vol.~58, no.~1, pp. 106--112, Jan. 2020.

\bibitem{Renzo2019Smart}
M.~Di~Renzo \emph{et~al.}, ``Smart radio environments empowered by
  reconfigurable {AI} meta-surfaces: An idea whose time has come,''
  \emph{EURASIP J. Wireless Commun. Netw.}, vol. 2019, no.~1, pp. 129--148, May
  2019.

\bibitem{zheng2021survey}
B.~Zheng, C.~You, W.~Mei, and R.~Zhang, ``A survey on channel estimation and
  practical passive beamforming design for intelligent reflecting surface aided
  wireless communications,'' \emph{IEEE Commun. Surveys Tuts.}, vol.~24, no.~2,
  pp. 1035--1071, Second Quarter 2022.

\bibitem{zheng2019intelligent}
B.~Zheng and R.~Zhang, ``Intelligent reflecting surface-enhanced {OFDM}:
  Channel estimation and reflection optimization,'' \emph{IEEE Wireless Commun.
  Lett.}, vol.~9, no.~4, pp. 518--522, Apr. 2020.

\bibitem{zheng2020intelligent}
B.~Zheng, C.~You, and R.~Zhang, ``Intelligent reflecting surface assisted
  multi-user {OFDMA}: Channel estimation and training design,'' \emph{IEEE
  Trans. Wireless Commun.}, vol.~19, no.~12, pp. 8315--8329, Dec. 2020.

\bibitem{yang2019intelligent}
Y.~Yang, B.~Zheng, S.~Zhang, and R.~Zhang, ``Intelligent reflecting surface
  meets {OFDM}: Protocol design and rate maximization,'' \emph{IEEE Trans.
  Commun.}, vol.~68, no.~7, pp. 4522--4535, Jul. 2020.

\bibitem{zheng2021irs}
B.~Zheng and R.~Zhang, ``{IRS} meets relaying: Joint resource allocation and
  passive beamforming optimization,'' \emph{IEEE Wireless Commun. Lett.},
  vol.~10, no.~9, pp. 2080--2084, Sept. 2021.

\bibitem{Yildirim2021Hybrid}
I.~{Yildirim}, F.~{Kilinc}, E.~{Basar}, and G.~C. {Alexandropoulos}, ``Hybrid
  {RIS}-empowered reflection and decode-and-forward relaying for coverage
  extension,'' \emph{IEEE Commun. Lett.}, vol.~25, no.~5, pp. 1692--1696, May
  2021.

\bibitem{Abdullah2021Optimization}
Z.~Abdullah, G.~Chen, S.~Lambotharan, and J.~A. Chambers, ``Optimization of
  intelligent reflecting surface assisted full-duplex relay networks,''
  \emph{IEEE Wireless Commun. Lett.}, vol.~10, no.~2, pp. 363--367, Feb. 2021.

\bibitem{Zheng2020IRSNOMA}
B.~{Zheng}, Q.~{Wu}, and R.~{Zhang}, ``Intelligent reflecting surface-assisted
  multiple access with user pairing: {NOMA or OMA}?'' \emph{IEEE Commun.
  Lett.}, vol.~24, no.~4, pp. 753--757, Apr. 2020.

\bibitem{Guo2021Intelligent}
Y.~Guo, Z.~Qin, Y.~Liu, and N.~Al-Dhahir, ``Intelligent reflecting surface
  aided multiple access over fading channels,'' \emph{IEEE Trans. Commun.},
  vol.~69, no.~3, pp. 2015--2027, Mar. 2021.

\bibitem{Zuo2021Reconfigurable}
J.~Zuo, Y.~Liu, and N.~Al-Dhahir, ``Reconfigurable intelligent surface assisted
  cooperative non-orthogonal multiple access systems,'' \emph{IEEE Trans.
  Commun.}, vol.~69, no.~10, pp. 6750--6764, Oct. 2021.

\bibitem{Lin2022Sensing}
S.~Lin, B.~Zheng, F.~Chen, and R.~Zhang, ``Intelligent reflecting surface-aided
  spectrum sensing for cognitive radio,'' \emph{IEEE Wireless Commun. Lett.},
  vol.~11, no.~5, pp. 928--932, May 2022.

\bibitem{buzzi2022foundations}
S.~Buzzi, E.~Grossi, M.~Lops, and L.~Venturino, ``Foundations of {MIMO} radar
  detection aided by reconfigurable intelligent surfaces,'' \emph{IEEE Trans.
  Signal Process.}, vol.~70, pp. 1749--1763, Mar. 2022.

\bibitem{Shao2022Target}
X.~Shao, C.~You, W.~Ma, X.~Chen, and R.~Zhang, ``Target sensing with
  intelligent reflecting surface: Architecture and performance,'' \emph{IEEE J.
  Sel. Areas Commun.}, vol.~40, no.~7, pp. 2070--2084, Jul. 2022.

\bibitem{wang2023target}
P.~Wang, W.~Mei, J.~Fang, and R.~Zhang, ``Target-mounted intelligent reflecting
  surface for joint location and orientation estimation,'' \emph{arXiv preprint
  arXiv:2301.09248}, 2023.

\bibitem{Wang2022Joint}
X.~Wang, Z.~Fei, J.~Huang, and H.~Yu, ``Joint waveform and discrete phase shift
  design for {RIS}-assisted integrated sensing and communication system under
  {Cramer-Rao} bound constraint,'' \emph{IEEE Trans. Veh. Technol.}, vol.~71,
  no.~1, pp. 1004--1009, Jan. 2022.

\bibitem{Hu2022Reconfigurable}
J.~Hu, H.~Zhang, B.~Di, L.~Li, K.~Bian, L.~Song, Y.~Li, Z.~Han, and H.~V. Poor,
  ``Reconfigurable intelligent surface based {RF} sensing: Design,
  optimization, and implementation,'' \emph{IEEE J. Sel. Areas Commun.},
  vol.~38, no.~11, pp. 2700--2716, Nov. 2020.

\bibitem{Wu2019TWC}
Q.~Wu and R.~Zhang, ``Intelligent reflecting surface enhanced wireless network
  via joint active and passive beamforming,'' \emph{IEEE Trans. Wireless
  Commun.}, vol.~18, no.~11, pp. 5394--5409, Nov. 2019.

\bibitem{Zheng2020DoubleIRS}
B.~Zheng, C.~You, and R.~Zhang, ``Double-{IRS} assisted multi-user {MIMO}:
  Cooperative passive beamforming design,'' \emph{IEEE Trans. Wireless
  Commun.}, vol.~20, no.~7, pp. 4513--4526, Jul. 2021.

\bibitem{zheng2020efficient}
------, ``Efficient channel estimation for double-{IRS} aided multi-user {MIMO}
  system,'' \emph{IEEE Trans. Commun.}, vol.~69, no.~6, pp. 3818--3832, Jun.
  2021.

\bibitem{mei2021intelligent}
W.~Mei, B.~Zheng, C.~You, and R.~Zhang, ``Intelligent reflecting surface-aided
  wireless networks: From single-reflection to multi-reflection design and
  optimization,'' \emph{Proc. IEEE}, vol. 110, no.~9, pp. 1380--1400, May 2022.

\bibitem{huang2021Multi-Hop}
C.~Huang, Z.~Yang, G.~C. Alexandropoulos, K.~Xiong, L.~Wei, C.~Yuen, Z.~Zhang,
  and M.~Debbah, ``Multi-hop {RIS}-empowered terahertz communications: A
  {DRL}-based hybrid beamforming design,'' \emph{IEEE J. Sel. Areas Commun.},
  vol.~39, no.~6, pp. 1663--1677, Jun. 2021.

\bibitem{zheng2022intelligent}
B.~Zheng, S.~Lin, and R.~Zhang, ``Intelligent reflecting surface-aided {LEO}
  satellite communication: Cooperative passive beamforming and distributed
  channel estimation,'' \emph{IEEE J. Sel. Areas Commun.}, vol.~40, no.~10, pp.
  3057--3070, Oct. 2022.

\bibitem{wei2021channel}
L.~Wei, C.~Huang, G.~C. Alexandropoulos, C.~Yuen, Z.~Zhang, and M.~Debbah,
  ``Channel estimation for {RIS}-empowered multi-user {MISO} wireless
  communications,'' \emph{IEEE Trans. Commun.}, vol.~69, no.~6, pp. 4144--4157,
  Jun. 2021.

\bibitem{huang2020reconfigurable}
C.~Huang, R.~Mo, and C.~Yuen, ``Reconfigurable intelligent surface assisted
  multiuser {MISO} systems exploiting deep reinforcement learning,'' \emph{IEEE
  J. Sel. Areas Commun.}, vol.~38, no.~8, pp. 1839--1850, Aug. 2020.

\bibitem{zhang2019capacity}
S.~Zhang and R.~Zhang, ``Capacity characterization for intelligent reflecting
  surface aided {MIMO} communication,'' \emph{IEEE J. Sel. Areas Commun.},
  vol.~38, no.~8, pp. 1823--1838, Aug. 2020.

\bibitem{Pan2020Multicell}
C.~{Pan}, H.~{Ren}, K.~{Wang}, W.~{Xu}, M.~{Elkashlan}, A.~{Nallanathan}, and
  L.~{Hanzo}, ``Multicell {MIMO} communications relying on intelligent
  reflecting surfaces,'' \emph{IEEE Trans. Wireless Commun.}, vol.~19, no.~8,
  pp. 5218--5233, Aug. 2020.

\bibitem{Xie2021Max}
H.~Xie, J.~Xu, and Y.-F. Liu, ``Max-min fairness in {IRS}-aided multi-cell
  {MISO} systems with joint transmit and reflective beamforming,'' \emph{IEEE
  Trans. Wireless Commun.}, vol.~20, no.~2, pp. 1379--1393, Feb. 2021.

\bibitem{Luo2021Reconfigurable}
C.~Luo, X.~Li, S.~Jin, and Y.~Chen, ``Reconfigurable intelligent
  surface-assisted multi-cell {MISO} communication systems exploiting
  statistical {CSI},'' \emph{IEEE Wireless Commun. Lett.}, vol.~10, no.~10, pp.
  2313--2317, Oct. 2021.

\bibitem{Yu2022Robust}
X.~Yu, D.~Xu, Y.~Sun, D.~W.~K. Ng, and R.~Schober, ``Robust and secure wireless
  communications via intelligent reflecting surfaces,'' \emph{IEEE J. Sel.
  Areas Commun.}, vol.~38, no.~11, pp. 2637--2652, Nov. 2020.

\bibitem{Shen2019Secrecy}
H.~Shen, W.~Xu, S.~Gong, Z.~He, and C.~Zhao, ``Secrecy rate maximization for
  intelligent reflecting surface assisted multi-antenna communications,''
  \emph{IEEE Commun. Lett.}, vol.~23, no.~9, pp. 1488--1492, Sept. 2019.

\bibitem{Sheng2020Artificial}
S.~Hong, C.~Pan, H.~Ren, K.~Wang, and A.~Nallanathan, ``Artificial-noise-aided
  secure {MIMO} wireless communications via intelligent reflecting surface,''
  \emph{IEEE Trans. Commun.}, vol.~68, no.~12, pp. 7851--7866, Sept. 2020.

\bibitem{grant2014cvx}
\BIBentryALTinterwordspacing
M.~Grant and S.~Boyd, ``{CVX}: Matlab software for disciplined convex
  programming,'' 2016. [Online]. Available: \url{http://cvxr.com/cvx}
\BIBentrySTDinterwordspacing

\bibitem{Luo2010Semidefinite}
Z.~{Luo}, W.~{Ma}, A.~M. {So}, Y.~{Ye}, and S.~{Zhang}, ``Semidefinite
  relaxation of quadratic optimization problems,'' \emph{IEEE Signal Process.
  Mag.}, vol.~27, no.~3, pp. 20--34, May 2010.

\bibitem{Swindlehurst1992MUSIC}
A.~Swindlehurst and T.~Kailath, ``A performance analysis of subspace-based
  methods in the presence of model errors. {I}. {The MUSIC} algorithm,''
  \emph{IEEE Trans. Signal Process.}, vol.~40, no.~7, pp. 1758--1774, Jul.
  1992.

\end{thebibliography}

\end{document}